\documentclass[traditabstract]{aa}

\usepackage{amsmath}
\usepackage{graphicx}
\usepackage{xspace}
\usepackage{natbib}
\usepackage{txfonts}
\usepackage{hyperref}
\usepackage{rotating}

\bibpunct{(}{)}{;}{a}{}{,}

\newcommand{\suz}{\textsl{Suzaku}\xspace}

\newcommand{\nustar}{\textsl{NuSTAR}\xspace}
\newcommand{\integral}{\textsl{INTEGRAL}\xspace}
\newcommand{\rxte}{\textsl{RXTE}\xspace}
\newcommand{\swift}{\textsl{Swift}\xspace}
\newcommand{\fermi}{\textsl{Fermi}\xspace}
\newcommand{\gaia}{\textsl{Gaia}\xspace}
\newcommand{\bepposax}{\textsl{Beppo}SAX\xspace}

\newcommand{\cutoffpl}{\texttt{CutoffPL}\xspace}
\newcommand{\highecut}{\texttt{PLCUT}\xspace}
\newcommand{\npex}{\texttt{NPEX}\xspace}
\newcommand{\fdcut}{\texttt{FDCUT}\xspace}
\newcommand{\compmag}{\texttt{compmag}\xspace}
\newcommand{\tbnew}{\texttt{tbnew}\xspace}
\newcommand{\gabs}{\texttt{gabs}\xspace}

\newcommand{\gauss}{\texttt{Gauss}\xspace}

\newcommand{\ax}{\object{A~0535+26}\xspace}

\begin{document}
\title{Looking at \ax at low luminosities with \nustar}

\author{
     Ralf Ballhausen\inst{1} 
\and Katja Pottschmidt\inst{2,3} 
\and Felix F\"urst\inst{4,5} 
\and J\"orn Wilms\inst{1}
\and John A.\ Tomsick\inst{6}
\and Fritz-Walter Schwarm\inst{1}
\and Daniel Stern\inst{7}
\and Peter Kretschmar\inst{5}
\and Isabel Caballero\inst{1}
\and Fiona A.\ Harrison\inst{4}
\and Steven E.\ Boggs\inst{6}
\and Finn E.\ Christensen\inst{8}
\and William W.\ Craig\inst{6,9}
\and Charles J.\ Hailey\inst{10}
\and William W.\ Zhang\inst{11}
}

\authorrunning{R.\ Ballhausen et al.}

\institute{
	Dr.\ Karl-Remeis-Sternwarte and Erlangen Centre for
        Astroparticle Physics, Sternwartstr.~7, 96049 Bamberg, Germany 
\and 
	Department of Physics and Center for Space Science and Technology, 
        UMBC, Baltimore, MD 21250, USA
\and
	CRESST and NASA Goddard Space Flight Center, Astrophysics Science
        Division, Code 661, Greenbelt, MD 20771, USA
\and 
	Cahill Center for Astronomy and Astrophysics, California Institute 
        of Technology, Pasadena, CA 91125, USA
\and
        European Space Astronomy Centre (ESA/ESAC), Science
        Operations Department, Villanueva de la Ca\~nada (Madrid), Spain
\and
        Space  Sciences  Laboratory,  7  Gauss  Way,  University  of 
        California, Berkeley, CA 94720-7450, USA
\and 
        Jet Propulsion Laboratory, California Institute of Technology, 
        Pasadena, CA 91109, USA
\and
        DTU Space, National Space Institute, Technical University of Denmark, 
        Elektrovej 327, DK-2800 Lyngby, Denmark 
\and    
        Lawrence Livermore National Laboratory, Livermore, CA 94550, USA
\and
        Columbia Astrophysics Laboratory, Columbia University, New York, 
        NY 10027, USA
\and
        NASA Goddard Space Flight Center, Greenbelt, MD 20771, USA
}

\abstract{We report on two \nustar observations of the high-mass X-ray
  binary \ax taken toward the end of its normal 2015 outburst at very
  low 3--50\,keV luminosities of
  ${\sim}1.4\times10^{36}\,\mathrm{erg}\,\mathrm{s}^{-1}$ and
  ${\sim}5\times10^{35}\,\mathrm{erg}\,\mathrm{s}^{-1}$, which are
  complemented by nine \swift observations. The data clearly confirm
  indications seen in earlier data that the source's spectral shape
  softens as it becomes fainter. The smooth, exponential rollover at
  high energies present in the first observation evolves to a much
  more abrupt steepening of the spectrum at 20--30\,keV. The continuum
  evolution can be well described with emission from a magnetized
  accretion column, modeled using the \compmag model modified by an
  additional Gaussian emission component for the fainter
  observation. Between the two observations, the optical depth changes
  from $0.75\pm0.04$ to $0.56^{+0.01}_{-0.04}$, the electron
  temperature remains constant, and there is an indication that the
  column decreases in radius.  Since the energy resolved pulse
  profiles remain virtually unchanged in shape between the two
  observations, the emission properties of the accretion column,
  however, reflect the same accretion regime. This conclusion is also
  confirmed by our result that the energy of the cyclotron resonant
  scattering feature (CRSF) at $\sim$45\,keV is independent of the
  luminosity, implying that the magnetic field in the region in which
  the observed radiation is produced is the same in both
  observations. Finally, we also constrain the evolution of the
  continuum parameters with rotational phase of the neutron star. The
  width of the CRSF could only be constrained for the brighter
  observation. Based on Monte-Carlo simulations of CRSF formation in
  single accretion columns, its pulse phase dependence supports a
  simplified fan beam emission pattern. The evolution of the CRSF
  width is very similar to that of the CRSF depth, which is, however,
  in disagreement with expectations.}

\date{Received DATE / Accepted DATE} 

\maketitle

\section{Introduction}

The Be/X-ray binary \ax was discovered during routine observations of
the Crab with the Rotation Modulation Collimator on \textsl{Ariel V}
as a transient X-ray pulsar with a period of 104\,s
\citep{Rosenberg1975}. Its optical companion HDE~245770 \citep{Li1979}
is a B0IIIe star with an estimated distance of $\sim$2\,kpc
\citep{Steele1998}. This distance estimate has been confirmed by the
first \gaia data release, which also reports a parallax corresponding
to a distance of $\sim$2\,kpc\footnote{The relative uncertainty of the
  parallax of \ax in \gaia DR1 is $\sim$0.5. For such large
  uncertainties, the uncertainty of the distance estimate from
  parallax measurements has a systematic biased \citep{Lutz1973} that
  does not allow to assign an uncertainty to a single parallax derived
  distance.} \citep{Gaia2016a}. The orbital period and eccentricity
were measured to be $P_\mathrm{orb} = 111.1\pm0.3$\,d and $e =
0.42\pm0.02$, respectively \citep{Finger1996}.

\ax shows both type~I and~II outbursts. Type~I outbursts are
associated with the periastron passage of the neutron star, which
results in enhanced mass transfer from the donor star. In contrast,
type~II outbursts may occur at any orbital phase and are likely caused
by varying activity of the donor. A disk truncation mechanism able to
produce these different kinds of outbursts is discussed by
\citet{Okazaki2001}. Intervals with very regular type I outbursts at
each periastron passage are interrupted by periods of quiescence,
sometimes lasting several years. The peak fluxes of the outbursts can
vary by several orders of magnitude. They are normally around a few
hundred mCrab, but reached $\sim$8\,Crab in the 20--40\,keV band in
1994 \citep{Wilson1994}. The outburst studied here was of medium
strength relative to the history of this source, with a peak
15--50\,keV flux of $\sim$600\,mCrab in \swift/BAT. The X-ray pulsar
and the optical companion \object{HDE 245770} has been monitored
extensively. A correlation of X-ray and optical activity is discussed
by, e.g., \citet{Camero-Arranz2012}, who report the presence of an
H$\alpha$ line mostly in emission, indicating a persistent but
variable Be disk. Episodes of increased optical brightness and a
strong H$\alpha$ emission line were found to precede high X-ray
activity. The correlation of V magnitude and H$\alpha$ equivalent
width changed in long-term observations from 1992 to 2010. This
behavior might be connected to a renewal of the Be disk and mass
ejections of the donor star \citep{Yan2012}.

The X-ray spectrum of \ax has often been successfully modeled with a
cutoff powerlaw with an additional black body component with a
temperature of 1--2\,keV \citep[e.g.,][]{Caballero2013}. \ax is a
known cyclotron resonant scattering feature (CRSF or cyclotron line)
source with a fundamental line energy around 50\,keV and a second
harmonic around 100\,keV
\citep{Kendziorra1994,Grove1995,Kretschmar1996}. Cyclotron lines in
the spectra of accreting X-ray pulsars result from the inelastic
scattering of photons off electrons in a strong magnetic field, where
the electron momenta perpendicular to the magnetic field are quantized
\citep[e.g.,][and references
  therein]{Canuto1977,Sina1996,Schoenherr2007,Schwarm2016a}. Measuring
their energy allows us to probe the magnetic field at the location
where the observed radiation is produced.

For a long time, observations appeared to show that the energy of the
cyclotron line was independent of luminosity, indicating that the
region in which the line is formed in the accretion column is at a
stable location in the neutron star's magnetic field
\citep{Caballero2007, Caballero2013}. Using pulse amplitude resolved
spectroscopy, however, \citet{Klochkov2011}, found a positive
correlation of the cyclotron energy with luminosity, while at very
high luminosities, above $10^{37}\,\mathrm{erg}\,\mathrm{s}^{-1}$,
pulse-averaged spectroscopy also shows that the cyclotron line energy
depends on luminosity \citep{Sartore2015}. Such a behavior is expected
from radiative shock models for the accretion column, where the
location of the shock depends on the luminosity \citep{Becker2012}.

With an estimated distance of only $\sim$2\,kpc, \ax is a good
candidate to study accretion mechanisms at very low luminosities.
\citet{Rothschild2013} present several \rxte observations close to
quiescence, although pulsations were present in some of the
observations. During this phase, the X-ray spectrum is well described
by either a powerlaw or a bremsstrahlung model.

Another series of quiescence observations were taken by \bepposax in
2000/2001 at luminosities of
${\sim}1.5\times10^{33}\,\mathrm{erg}\,\mathrm{s}^{-1}$ and
${\sim}4.4\times10^{33}\,\mathrm{erg}\,\mathrm{s}^{-1}$ in 2--10\,keV
\citep{Orlandini2004}. The source showed pulsations which indicate
that accretion taking place at a luminosity where mass transfer should
be inhibited by the centrifugal barrier (so-called propeller
regime). The spectral shape could be described by an absorbed power
law as well as a thermal bremsstrahlung model. Further, these authors
found evidence for the second harmonic CRSF at ${\sim}118$\,keV
although the fundamental could not be detected.

Here, we study two \nustar and nine \swift observations of \ax
performed toward the end of the 2015 outburst, which complement these
earlier results. The remainder of the paper is organized as follows:
In Sect.~\ref{sec:data_reduction} we describe the data acquisition and
reduction process. Section~\ref{sec:timing} is dedicated to the pulse
profiles and their energy dependence. In
Sect.~\ref{sec:spectroscopy_phavg} and \ref{sect:spectroscopy_phres}
we present phase-averaged and phase-resolved spectroscopy,
respectively, and in Sect.~\ref{sec:discussion} we summarize and
discuss our results.

\section{Data acquisition and reduction}\label{sec:data_reduction}

\begin{figure}
\resizebox{\hsize}{!}{\includegraphics{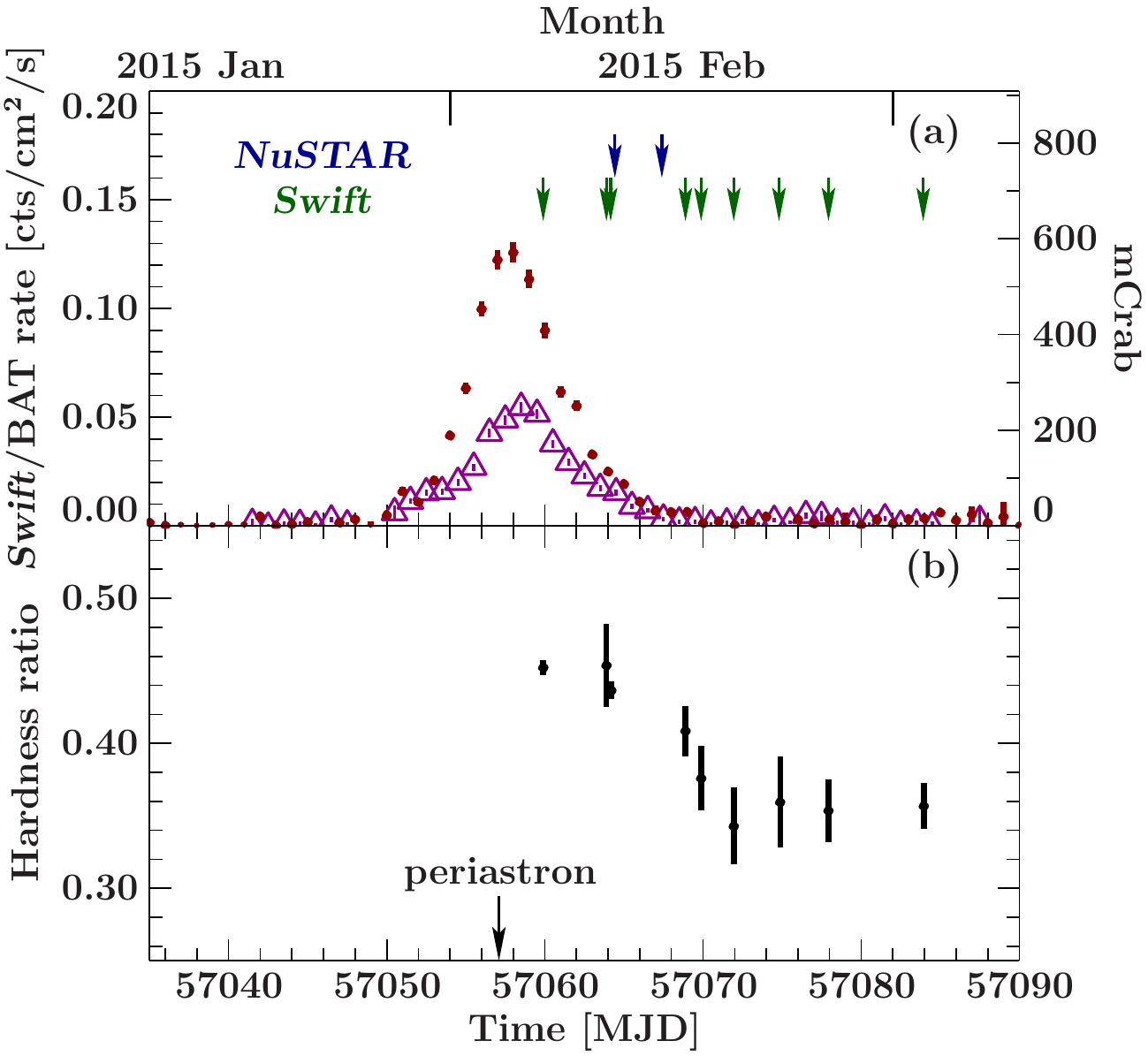}}
\caption{\textbf{(a)} Brown points show the \swift/BAT (15--50\,keV)
  daily lightcurve \citep{Krimm2013} with times of \nustar and pointed
  \swift/XRT observations marked (arrows). Purple triangles show the
  MAXI/GSC \citep{Mihara2011} daily lightcurve (4--10\,keV), rescaled
  to mCrab fluxes (right-hand $y$-axis). \textbf{(b)} Hardness ratio
  of \swift/XRT observations, defined as the count rate of the
  4--7\,keV band divided by the count rate of the 1--4\,keV band.}
\label{fig:bat_lc}
\end{figure}

Figure~\ref{fig:bat_lc} shows the \swift/BAT and MAXI/GSC daily
lightcurves and the \swift/XRT derived evolution of the hardness ratio
of the 2015 outburst of \ax. The times of the \nustar and \swift/XRT
observations are marked.

\subsection{\swift}

NASA's \swift satellite \citep{Gehrels2004, Gehrels2005} was launched
as a Medium Explorer mission in 2004. Its scientific payload
consists of the \textsl{Burst Alert Telescope}
\citep[BAT,][]{Barthelmy2005}, the \textsl{X-ray Telescope}
\citep[XRT,][]{Burrows2005}, and the \textsl{Ultraviolet/Optical
  Telescope} \citep[UVOT,][]{Roming2005}. Besides conducting a
continuous all-sky survey in the hard X-rays with the BAT, \swift is
also capable of pointed observations with higher spectral and spatial
resolution.

The XRT uses a grazing incidence Wolter type~I telescope and a charge
coupled device (CCD) sensitive in the energy range of 0.2--10\,keV.
The XRT operates in different observing modes, resulting in different
read-out times, which are appropriate for different flux levels. The
effective area of the XRT is more than 120\,$\mathrm{cm}^2$ at
1.5\,keV.

\swift/XRT took ten $\sim$1\,ks snapshot observations of \ax in
February and March 2015 (Table~\ref{tab:obs_log}). All observations
were taken in \textsl{Windowed Timing Mode}, which provides a time
resolution of 1.7\,ms, but only one spatial dimension. We discarded
ObsID~00035066058 because the source could not be localized on the
chip. We used HEAsoft v.\,6.18 and CalDB v.\,20160121 for data
reprocessing and extraction. Source and background regions are stripes
of $96''$ width, centered at the source position and at the outer
areas of the chip, respectively. None of the observations was affected
by pile-up.

\subsection{\nustar}

The \textsl{Nuclear Spectroscopic Telescope Array}
\citep[\nustar,][]{Harrison2013} was launched on 2012 June 13 as a
NASA Small Explorer mission. It carries two co-aligned grazing
incidence X-ray telescopes and solid-state detectors, sensitive to the
energy range of 3--79\,keV.

The optics are built of 133 nested shells with a field of view of
$10'\times10'$ at 10\,keV. The detectors placed in the focal plane of
each optical module (called FPM-A and FPM-B) are pixeled CdZnTe
detectors which are actively shielded by CsI detectors. The pixels are
read out individually upon triggering, which avoids pile-up, a severe
problem in many other imaging X-ray detectors employing CCDs. Incident
count rates of
${\sim}10^5\,\mathrm{cts}\,\mathrm{s}^{-1}\,\mathrm{pixel}^{-1}$ can
in principle be observed without significant pile-up
\citep{Harrison2013}, although in practice the maximum events rate
that can be detected in a module is limited to
${\sim}400\,\mathrm{s}^{-1}$ due to to the read-out and processing
time of each event. The time resolution is $2\,\mu\mathrm{s}$
\citep{Harrison2013, Bachetti2015}.

\ax was observed by \nustar on 2015 February 11 and 2015 February 14
(ObsID 80001016002 and 80001016004, hereafter Obs.~I and Obs.~II,
respectively). The data were reprocessed and extracted using the
standard \texttt{NUSTARDAS} pipeline v.\,1.6 with CalDB version
20161021. Source and background regions are circles with $90''$
radius. All timing information was transferred to the solar barycenter
with the FTOOL \texttt{barycorr} and corrected for binary motion
according to the ephemeris of \citet{Finger1996} with updated orbital
period and epoch\footnote{An updated period is available through the
  \fermi/GBM pulsar page
  \url{https://gammaray.msfc.nasa.gov/gbm/science/pulsars/lightcurves/a0535.html}}.

\begin{table}
  \caption{Observation log of the \nustar and \swift/XRT observations.} 
\label{tab:obs_log}
\centering
\begin{tabular}{ccc}
\hline
ObsID & mid-time~[MJD] & exposure~[ks] \\
\hline\hline
\multicolumn{3}{c}{\nustar} \\
\hline
80001016002 & 57064.43 & 21.4 \\
80001016004 & 57067.40 & 29.7 \\
\hline
\multicolumn{3}{c}{\swift} \\
\hline
00035066050 & 57059.91 & 1.09 \\
00035066051 & 57063.91 & 0.95 \\
00081432001 & 57064.21 & 1.94 \\
00035066052 & 57068.90 & 1.08 \\
00035066053 & 57069.90 & 1.06 \\
00035066054 & 57071.95 & 1.07 \\
00035066055 & 57074.88 & 0.71 \\
00035066056 & 57077.94 & 0.96 \\
00035066057 & 57083.96 & 1.55 \\
\hline
\end{tabular}
\end{table}

\section{Timing analysis}\label{sec:timing}

\begin{figure}
\centering
\resizebox{\hsize}{!}{\includegraphics{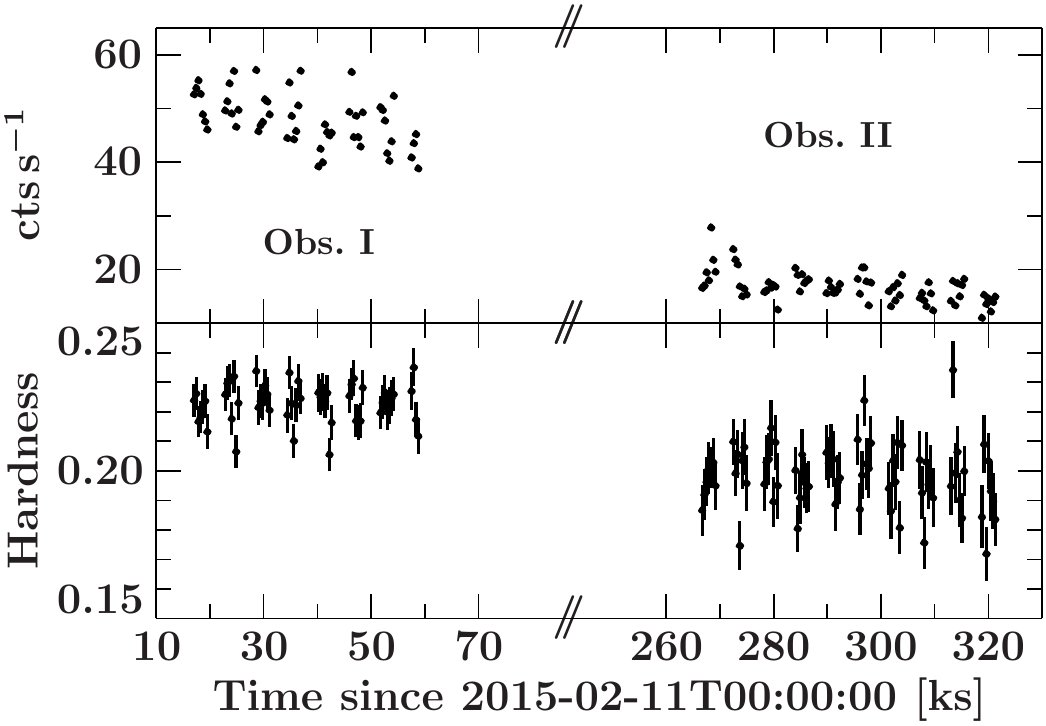}}
\caption{Background-subtracted lightcurve and hardness ratio for
  \nustar/FPMA data from both observations of \ax. To reduce
  pulse-phase variations, the time resolution of 413.6\,s was chosen
  to be an integer multiple of the mean pulse period. The hardness
  ratio is defined as the ratio of the count rate in the 15--50\,keV
  band divided by that of the 3--10\,keV band.}
\label{fig:lc_hardness}
\end{figure}

We extracted \nustar lightcurves with 413.6\,s time resolution for the
full 3--78\,keV range. In order to avoid variability only due to
pulsations, the binsize was chosen to be four times the mean pulse
period. The lightcurves are shown in Fig.~\ref{fig:lc_hardness},
together with the hardness ratio, defined here as the count rate of
the 15--50\,keV band divided by the count rate of the 3--10\,keV band.
The decrease in count rate over time is clearly visible and Obs.~II is
softer, in agreement with the long-term trend of the hardness ratio
evolution shown in Fig~\ref{fig:bat_lc}. Both lightcurves also show
moderate variability, which might be due to the well-known strong
pulse-to-pulse variability, \citep[e.g.,][]{Frontera1985,
  Klochkov2011}.

We determined the pulse period in both \nustar observations using the
epoch folding technique \citep{Leahy1983} applied to 0.5\,s resolved
lightcurves. The pulse periods during the \nustar observations are
103.3913(8)\,s in Obs.~I and 103.3890(9)\,s (both at 68\% confidence
level) in Obs.~II. Uncertainties on the pulse period were calculated
of a set of simulated lightcurves which are based on the previously
determined pulse period and profile with additional Gaussian noise.
The pulse period changes only very slightly between the observations
and is in excellent agreement with the pulse periods measured around
these times by \fermi/GBM \citep{Finger2009}. The \swift observations
are too short for constraining the pulse period.

\begin{figure}
\centering
\resizebox{\hsize}{!}{\includegraphics{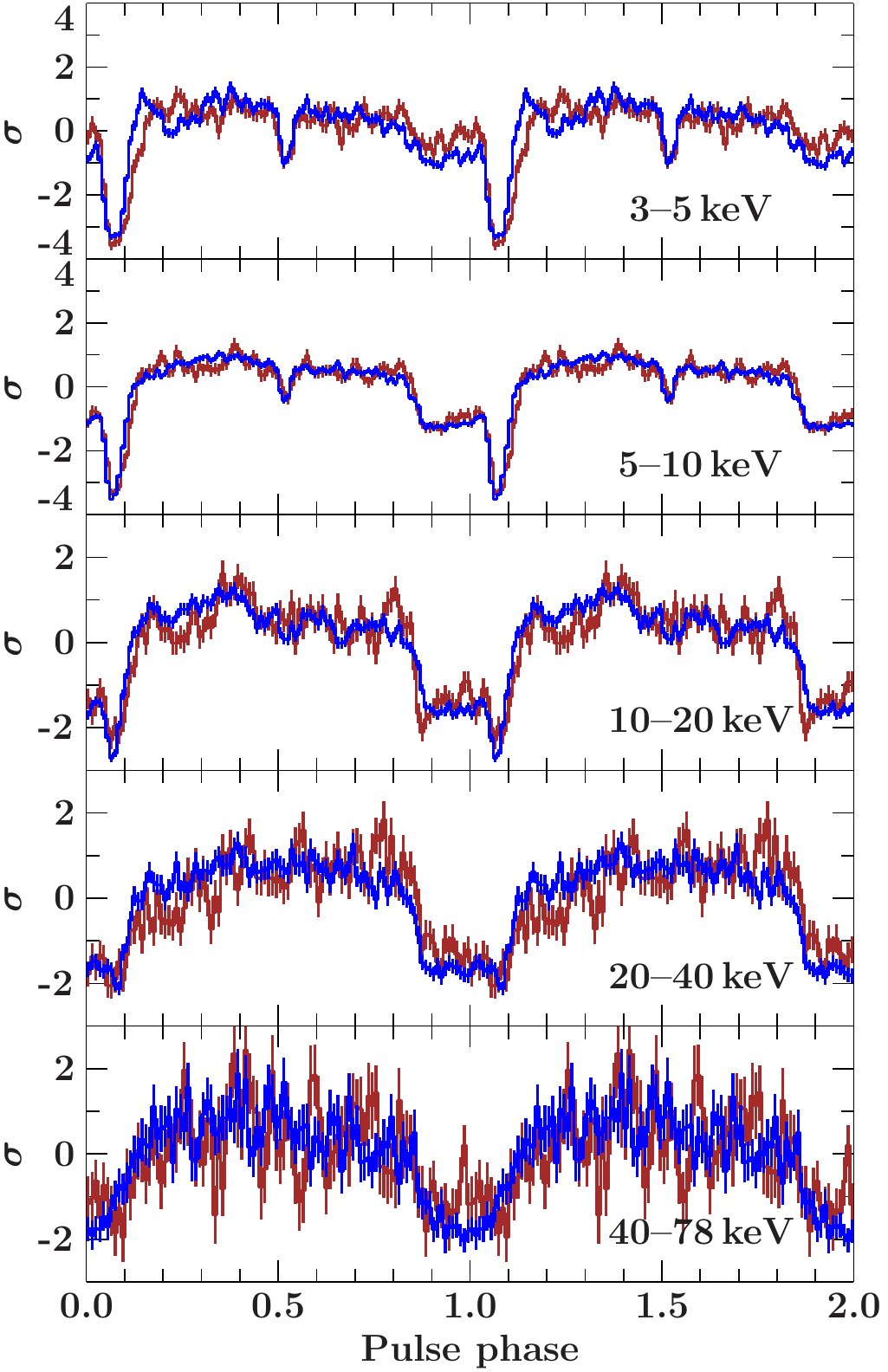}}
\caption{Energy-resolved, background-subtracted pulse profiles of
  \nustar/FPMA for Obs.~I (blue) and Obs.~II (red). All profiles are
  normalized such that their mean value is zero and their standard
  deviation is unity. The pulse profile is repeated for clarity.}
\label{fig:pulseprofiles}
\end{figure}

We folded the lightcurve on the local pulse period to obtain the pulse
profiles (Fig.~\ref{fig:pulseprofiles}). The pulse profiles have been
aligned by eye at the pulse minimum. The energy-resolved pulse
profiles are very similar in both observations, indicating the same
accretion geometry. They show an evolution from a broad, plateau-like
peak with several very narrow dips at lower energies to a rather
smooth, symmetric shape at higher energies.

\section{Spectral analysis}\label{sec:spectroscopy_phavg}

Traditionally, continua of accretion-powered X-ray pulsars are
described using empirical models which often consist of some powerlaw
component with a high-energy cutoff and sometimes an additional soft
component. A detailed description of the different empirical models of
the exponential turn-over and the terminology used here is given by
\citet{Mueller2013}.

More sophisticated, physical continuum models aim at calculating the
spectral shape by solving the radiative transfer equation for photons
passing through the accretion column. Such photons can be generated,
e.g., by bremsstrahlung or black body emission and are then modified
by Compton scattering with the electron plasma of the infalling
matter.  Recently, several physical model implementations have become
available. These models rely on slightly different assumptions on the
accretion geometry, velocity profile, and emission processes, and also
use different techniques to solve the radiation transfer problem.
Examples are the \compmag model \citep[][see Sect.~\ref{seq:compmag}
for more details]{Farinelli2012} and the Becker \& Wolff model
\citep[\texttt{BWmod};][]{Becker2007}.

In our analysis, we will first apply a sample of different empirical
continuum models and compare their characteristics and discuss their
success in describing the low luminosity observations. Then we test a
physical continuum model on the data, suitable for the low-luminosity
observations reported here. For all fits we used the 1--7\,keV and
3.5--79\,keV spectra of \swift/XRT and \nustar, respectively. We
jointly fitted \swift ObsID 00081432001 with \nustar Obs.~I and \swift
ObsID 00035066052 with \nustar Obs.~II. We rebinned the spectra of
FPMA and FPMB jointly, ensuring a minimum signal-to-noise ratio (SNR)
of 15 and adding at least 2 and 4 bins together for 3.5--40\,keV and
40--79\,keV, respectively, in both observations. \swift/XRT spectra
were rebinned to a minimum SNR of 15 and 5 for Obs.~I and II,
respectively. In all fits, detector constants normalized to
\nustar/FMPA were introduced to account for flux cross-calibration
uncertainties between the instruments. Photoelectric absorption is
accounted for with the \tbnew model, an updated version of
\texttt{tbabs}\footnote{see
  \url{http://pulsar.sternwarte.uni-erlangen.de/wilms/research/tbabs/}},
with abundances and cross sections set according to \citet{Wilms2000}
and \citet{Verner1996}, respectively.

\subsection{Empirical models}

The \highecut model consists of a single powerlaw with photon index
$\Gamma$ and a high-energy cutoff. The parameters determining the
cutoff are the folding energy, $E_\mathrm{fold}$, and the cutoff
energy, $E_\mathrm{cut}$. From the cutoff energy onwards, an
exponential decrease in flux is applied on scales of the folding
energy,
\begin{equation}
  \highecut(E)=E^{-\Gamma}\left\{\begin{array}{ll} 1, & \mbox{where $E
    \le E_\mathrm{cut}$} \\ \exp \left(-\frac{E -
    E_\mathrm{cut}}{E_\mathrm{fold}} \right) , & \mbox{where $E>
    E_\mathrm{cut}$} \end{array}\right.
\end{equation}
The \cutoffpl model is a special case for $E_\mathrm{cut} = 0$. The
\highecut model is not continuously differentiable at the cutoff
energy, which can lead to line-like residuals in a spectral fit.
Therefore, great caution has to be taken if, e.g., additional spectral
components such as a CRSF are located close to the cutoff energy. To
avoid this issue, the \fdcut cutoff provides a smoother turnover at
the cutoff energy. Alternatively, the \npex model consists of two
\cutoffpl models with equal folding energy. The second powerlaw has a
\emph{negative} photon index\citep[][and references
  therein]{Makishima1999,Mueller2013}.

\subsection{Physical models}\label{seq:compmag}
 
In order to describe the spectrum with more physically motivated
models, we use the \texttt{BWmod} \citep{Becker2007} and the \compmag
model of \citet{Farinelli2012}.

\texttt{BWmod} is based a solution of the radiative transfer for a
specific velocity profile that is linear in the optical depth
\citep{Becker2007}. This assumption allows an analytical solution and
is well justified for high mass accretion rates, where a radiative
dominated shock is present. This model was successfully applied to the
spectrum of \object{Her X-1} \citet{Wolff2016}, for observations at
higher luminosity than that of \ax. As expected, our attempts to fit
\texttt{BWmod} to the low luminosity \ax data failed.  Statistically
acceptable fits could be achieved only for the first observation,
probably due to the higher number of parameters compared to the
empirical models. However, the fits produced parameter combinations
that violated underlying assumptions of the model.

The \compmag model is better suited for lower luminosity observations.
It allows for different velocity profiles, characterized by an index,
$\eta$, and a terminal velocity, $\beta_0$, which can be different
from zero \citet{Farinelli2012}. The model has been included in
\texttt{XSPEC} releases since version 12.8.0. Contrary to
\texttt{BWmod} the radiative transfer equation inside the column is
solved numerically. While a recent update \citep{Farinelli2016} adds
bremsstrahlung and cyclotron emission as sources for seed photons, we
used the 2012 version of the \compmag model \citep{Farinelli2012},
where all seed photons are caused by black body radiation, with the
intention to test a less complex model and because the low luminosity
observations do not necessarily justify the assumption of a radiative
dominated shock.

\subsection{The CRSF}

\ax is a well-established CRSF source and thus we included an
absorption line-like component in our model.

CRSFs are typically modeled by Gaussian optical depth line profiles
(called \gabs in ISIS/XSPEC) or pseudo-Lorentzian profiles
\citep[\texttt{cyclabs};][]{Mihara1990, Makishima1990}. It should be
noted that the width, depth, and line energy are different in the two
models \citep[i.e, in the \texttt{cyclabs} model, the line energy
  $E_\mathrm{CRSF}$ does not represent where the line is deepest;
  see][for a comparison of CRSF energies obtained with different
  models]{Staubert2014}.

We found that both models provide a satisfactory description of the
CRSF feature. We use \gabs for the rest of the analysis because of its
simplicity and because it has been used in most previous analyses of
this source, allowing our results to be directly comparable.

\subsection{Results of spectral modeling}

\begin{figure}
\centering
\resizebox{\hsize}{!}{\includegraphics{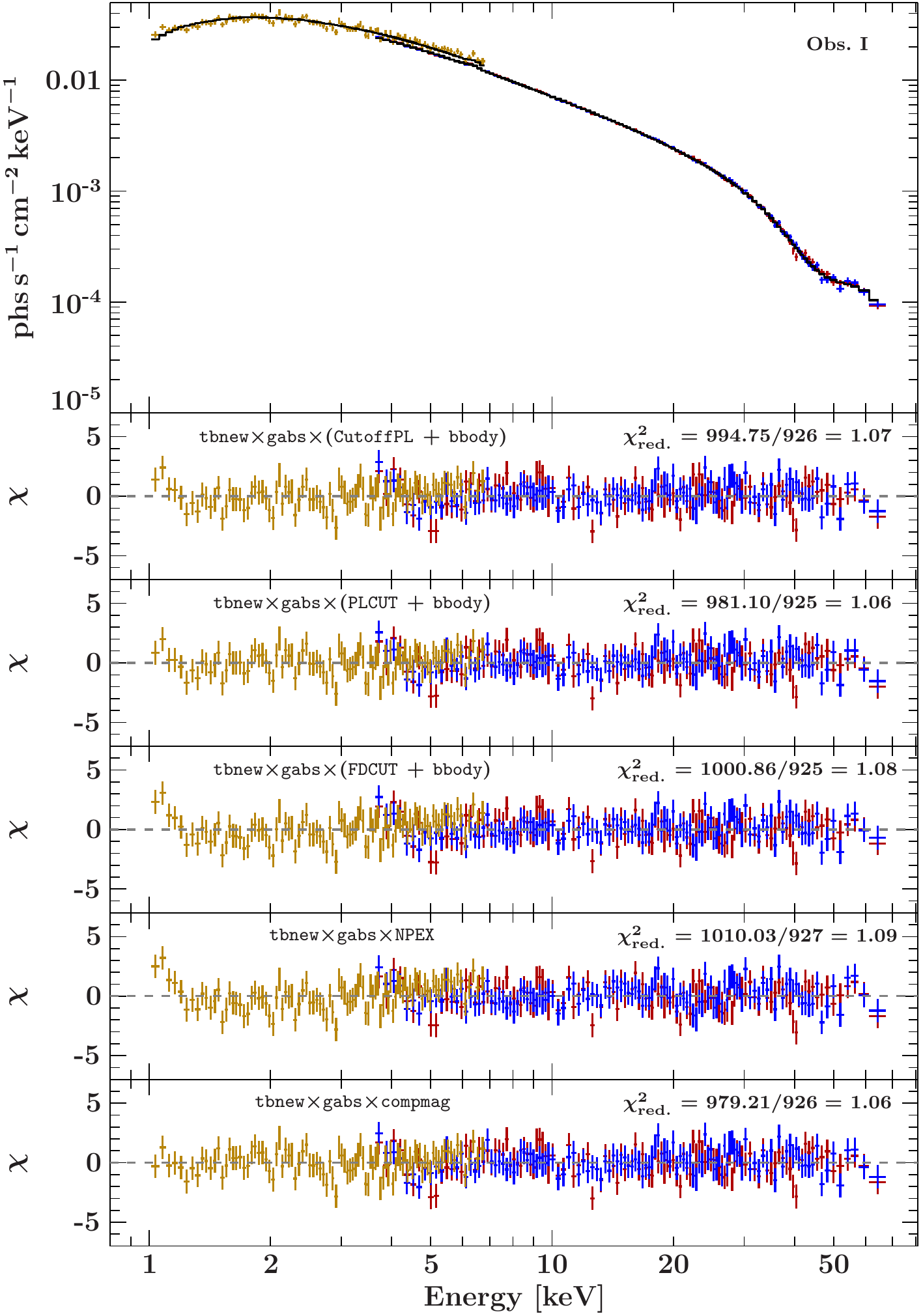}}
\caption{\textsl{Top panel:} Unfolded phase-averaged spectrum of
  Obs. I with best-fit model: XRT (gold), FPMA (blue), FPMB (red), and
  model for the \cutoffpl model (black). All models include an
  additional, narrow iron line. \textsl{Lower panels:} Residuals to
  the different continuum models. For clarity we binned the spectra
  using larger bins for the plot than the ones used for the fit.}
\label{fig:spectrum_ave_obs1}
\end{figure}

\begin{figure}
\centering
\resizebox{\hsize}{!}{\includegraphics{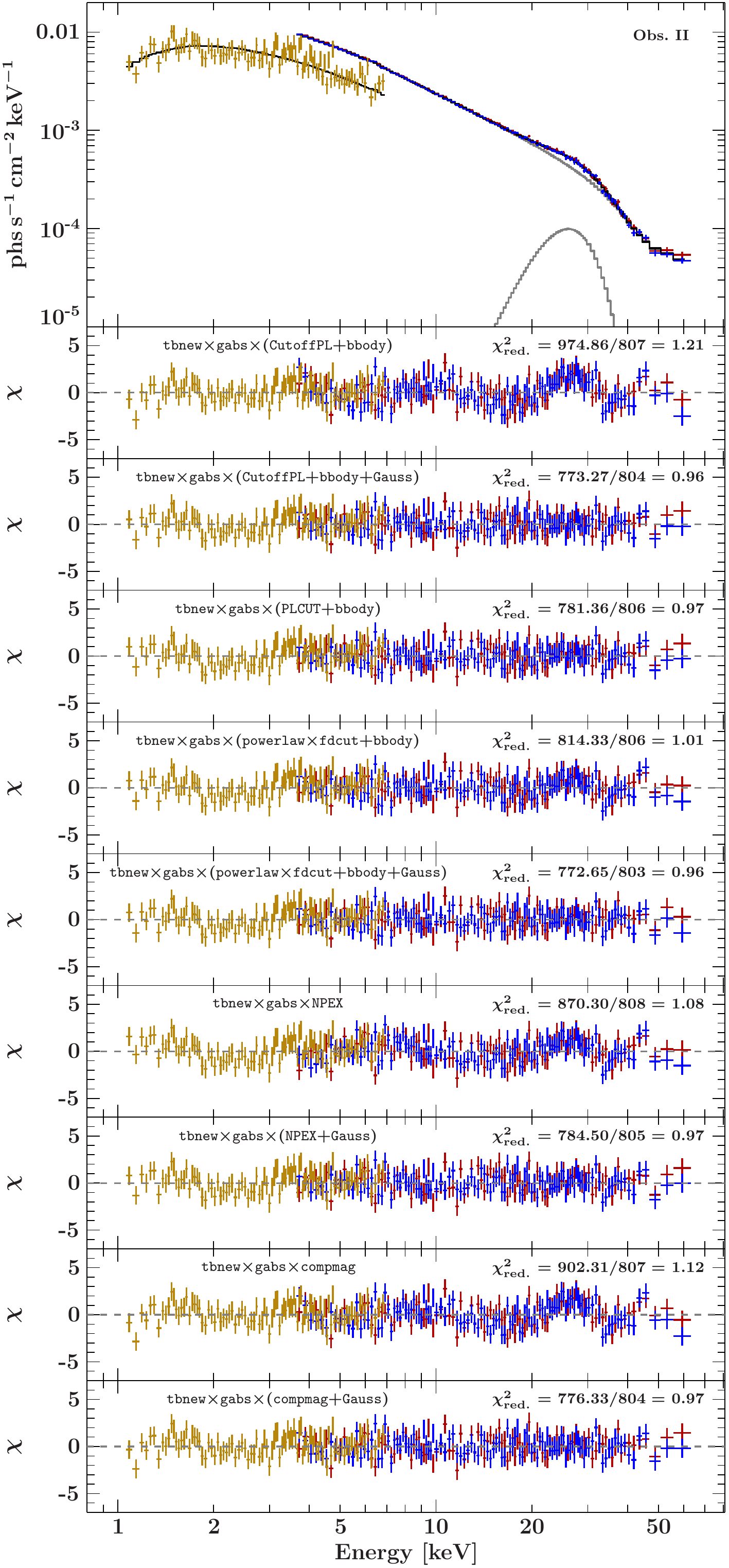}}
\caption{\textsl{Top panel:} Unfolded phase-averaged spectrum of
  Obs. II with the best-fit model: XRT (gold), FPMA (blue), FPMB
  (red), and model (black). All models include an additional, narrow
  iron line. The gray line shows a decomposition of the \cutoffpl and
  the Gaussian component. \textsl{Lower panels:} Residuals to the
  different continuum models. For clarity we binned the spectra using
  larger bins for the plot than the ones used for the fit.}
\label{fig:spectrum_ave_obs2}
\end{figure}

We tested the \cutoffpl, \highecut, \fdcut, \npex and \compmag models.
Best fit parameters are given for empirical models and both
observations in Table~\ref{tab:bestfit_paramer}. The spectra with one
best-fit model and residuals for all applied models for Obs.~I and II
are shown in Fig.~\ref{fig:spectrum_ave_obs1} and
\ref{fig:spectrum_ave_obs2}, respectively. All fit models, except
\npex and \compmag, require an additional soft black body.
Furthermore, a Gaussian emission line with a width fixed to
$10^{-6}$\,keV was included to model the Fe~K$\alpha$ line, thus the
line width is only determined by the detector response
(${\sim400}$\,eV for \nustar; \citealt{Harrison2013}, and
${\sim140}$\,eV for \swift/XRT; \citealt{Burrows2005}). Previous
observations have already indicated the presence of a narrow
Fe~K$\alpha$ line \citep{Caballero2007, Caballero2013}. We estimated
the significance of the inclusion of the iron line with a Monte-Carlo
approach \citep[see][for details]{Protassov2002}. Spectra are
simulated based on the model without the iron line component and then
fitted both with and without the iron line component and the $\chi^2$
difference is compared to the observed one. If the simulated $\chi^2$
difference is larger than the observed one we count this as a
false-positive detection of the iron line. In both observations, we
did not find any false-positive detection in 10\,000 simulations, so
the significance is $>99.99\%$. For Obs.~I the largest simulated
$\chi^2$ difference is $15.5$ while the observed one is $65.4$. For
Obs.~II the largest simulated $\chi^2$ difference is $14.4$ while the
observed one is $16.6$.

\begin{sidewaystable*}
\setlength{\tabcolsep}{5pt}
  \caption{Best-fit parameters for several empirical models for both
    observations. The single \cutoffpl model for Obs.~II is not
    acceptable, but shown here for comparison with the \cutoffpl+\gauss
    model.} 
  \label{tab:bestfit_paramer}
  \centering
  \begin{tabular}{lccccccccc}
      \hline\hline
                                                     & \multicolumn{2}{c}{\cutoffpl}                             & \cutoffpl + \gauss          & \multicolumn{2}{c}{\highecut}                             & \fdcut                            & \fdcut + \gauss            & \npex                          & \npex + \gauss                \\%
                                                     & Obs.~I                        & Obs.~II                   &  Obs.~II                    & Obs.~I                     &  Obs.~II                     & Obs.~I                            & Obs.~II                    & Obs.~I                         &  Obs.~II                      \\%
  \hline												                                                                                                   
  $N_\mathrm{H}$\tablefootmark{a}                    & $0.92^{+0.08}_{-0.07}$        & $0.64\pm0.16$             & $1.20^{+0.21}_{-0.20}$      & $0.84\pm0.08$              & $1.54^{+0.17}_{-0.16}$       & $1.11\pm0.07$                     & $1.37\pm0.08$              & $1.08\pm0.05$                  & $1.19^{+0.20}_{-0.18}$        \\%
\hline
  $\Gamma$                                           & $0.79\pm0.04$                 & $0.82\pm0.08$             & $1.07\pm0.08$               & $0.81\pm0.04$              & $1.41\pm0.02$                & $0.98\pm0.03$                     & $1.29\pm0.01$              & $0.68\pm0.02$                  & $0.61^{+0.19}_{-0.16}$        \\%
  $E_\mathrm{cut}$\tablefootmark{b}                  & --                            & --                        & --                          & $3.6^{+0.3}_{-0.4}$        & $27.9^{+1.1}_{-0.8}$         & $\le5$                            & $67.2^{+2.0}_{-3.9}$       & --                             & --                            \\%
  $E_\mathrm{fold}$\tablefootmark{b}                 & $20.2^{+1.1}_{-1.0}$          & $25.7^{+3.1}_{-2.3}$      & $27.4^{+3.6}_{-2.6}$        & $20.7^{+1.1}_{-1.0}$       & $27.4^{+2.3}_{-2.0}$         & $19.5^{+0.9}_{-0.8}$              & $\le26.7$                  & $9.6\pm0.3$                    & $8.0^{+1.4}_{-1.1}$           \\%
  $\mathcal{F}_\mathrm{PL}$\tablefootmark{c}         & $2.878\pm0.020$               & $0.929\pm0.012$           & $0.942^{+0.013}_{-0.028}$   & $2.886^{+0.020}_{-0.108}$  & $0.977\pm0.010$              & $2.894\pm0.020$                   & $0.915\pm0.003$            & $1.911\pm0.031$                & $0.508^{+0.071}_{-0.020}$     \\%
  $\mathcal{F}_\mathrm{PL,2}$\tablefootmark{c}       & --                            & --                        & --                          & --                         & --                           & --                                & --                         & $1.075\pm0.032$                & $0.380^{+0.032}_{-0.069}$     \\%
\hline  												                               	                                                                   
  $kT$\tablefootmark{b}                              & $1.30^{+0.04}_{-0.05}$        & $1.42^{+0.03}_{-0.02}$    & $1.44^{+0.09}_{-0.06}$      & $1.29\pm0.05$              & $1.76\pm0.04$                & $1.46^{+0.06}_{-0.05}$            & $1.58\pm0.03$              & --                             & --                            \\%
  $\mathcal{F}_\mathrm{BB}$\tablefootmark{c}         & $0.101\pm0.016$               & $0.104\pm0.010$           & $0.054^{+0.013}_{-0.012}$   & $0.091\pm0.016$            & $0.074\pm0.008$              & $0.093\pm0.016$                   & $0.063\pm0.001$            & --                             & --                            \\%
\hline													                               	                                                                   
  $\mathcal{F}_\mathrm{Gauss}$\tablefootmark{c}      & --                            & --                        & $0.051^{+0.029}_{-0.012}$   & --                         & --                           & --                                & $0.071\pm0.003$            & --                             & $0.208^{+0.128}_{-0.080}$      \\%
  $E_\mathrm{Gauss}$\tablefootmark{b}                & --                            & --                        & $26.1^{+0.9}_{-0.7}$        & --                         & --                           & --                                & $26.9\pm0.3  $             & --                             & $24.2^{+1.3}_{-2.2}$          \\%
  $\sigma_\mathrm{Gauss}$\tablefootmark{b}           & --                            & --                        & $4.9^{+1.1}_{-0.8}$         & --                         & --                           & --                                & $5.3\pm0.3$                & --                             & $7.8^{+1.5}_{-1.1}$           \\%
  \hline 													                               	                                                                   
  $E_\mathrm{CRSF}$\tablefootmark{b}                 & $45.5\pm0.7$                  & $47.2^{+1.1}_{-0.9}$      & $46.3^{+1.4}_{-1.8}$        & $45.7\pm0.7$               & $45.3^{+1.0}_{-0.9}$         & $45.3^{+0.8}_{-0.7}$              & $53.0\pm0.2$               & $45.8\pm0.8$                   & $35.0^{+6.0}_{-4.0}$                \\%
  $\sigma_\mathrm{CRSF}$\tablefootmark{b}            & $7.9^{+0.7}_{-0.6}$           & $7.4^{+0.8}_{-0.6}$       & $7.6^{+3.1}_{-1.4}$         & $8.0^{+0.7}_{-0.6}$        & $7.4^{+0.9}_{-0.8}$          & $8.3\pm0.7$                       & $17.7\pm0.3$               & $10.6^{+0.8}_{-0.7}$           & $19.6^{+0.5}_{-5.1}$          \\%
  $d_\mathrm{CRSF}$                                  & $12.4^{+1.8}_{-1.5}$          & $14.8^{+2.8}_{-2.1}$      & $9.1^{+4.6}_{-2.3}$         & $12.7^{+1.8}_{-1.5}$       & $10.6^{+2.2}_{-1.9}$         & $13.1^{+2.1}_{-1.7}$              & $72.0\pm0.6$               & $23.5^{+3.8}_{-3.0}$           & $\left(1.1\pm0.6\right)\times10^{2}$   \\%
\hline													                               	                                                                   
  $E_\mathrm{Fe}$\tablefootmark{b}                   & $6.40^{+0.04}_{-0.08}$        & $6.44^{+0.09}_{-0.12}$    & $6.44^{+0.05}_{-0.12}$      & $6.40^{+0.04}_{-0.08}$     & $6.40^{+0.09}_{-0.08}$       & $6.40^{+0.01}_{-0.08}$            & $6.44^{+0.05}_{-0.12}$     & $6.38^{+0.03}_{-0.06}$         & $6.38^{+0.07}_{-0.06}$        \\%
  $\mathrm{EQW}_\mathrm{Fe}$\tablefootmark{d}        & $20\pm5$                      & $8\pm6$                   & $14\pm6$                    & $21\pm5$                   & $13\pm6$                     & $19\pm4$                          & $13\pm5$                   & $20\pm4$                       & $17\pm6$                      \\%
\hline 													                               	                                                                   
  $c_\mathrm{FMPA}$\tablefootmark{e}                 & $1$                           & $1$                       & $1$                         & $1$                        & $1$                          & $1$                               & $1$                        & $1$                            & $1$                           \\%
  $c_\mathrm{FMPB}$\tablefootmark{e}                 & $1.005\pm0.003$               & $1.018\pm0.005$           & $1.018\pm0.005$             & $1.005\pm0.003$            & $1.018\pm0.005$              & $1.005\pm0.003$                   & $1.018\pm0.003$            & $1.005\pm0.003$                & $1.018\pm0.005$               \\%
  $c_\mathrm{XRT}$\tablefootmark{e}                  & $1.117\pm0.016$               & $0.545^{+0.023}_{-0.022}$ & $0.530\pm0.022$             & $1.127\pm0.017$            & $0.520\pm0.022$              & $1.105\pm0.016$                   & $0.525\pm0.018$            & $1.099\pm0.016$                & $0.528\pm0.022$               \\%
  \hline												                               	                                                                   
  $\chi^2_\mathrm{red}$ (d.o.f)                      & 1.07 (926)                    & 1.21 (807)                & 0.96 (804)                  & 1.06 (925)                 & 0.97 (807)                   & 1.08 (925)                        & 0.96 (803)                 & 1.09 (927)                     &  0.97 (805)   \\%
  \hline
\end{tabular}
\tablefoot{
  \tablefoottext{a}{Equivalent hydrogen column density in units of $10^{22}\,\mathrm{cm}^{-2}$.}
  \tablefoottext{b}{In keV.}
  \tablefoottext{c}{3--50\,keV flux in units of $10^{-9}\,\mathrm{erg}\,\mathrm{s}^{-1}\,\mathrm{cm}^{-2}$.}
  \tablefoottext{d}{Equivalent width in units of eV.}
  \tablefoottext{e}{Detector cross-calibration constant with respect to FPMA.}
}
\end{sidewaystable*}

For Obs.~I all four empirical continuum models result in statistically
acceptable fits. While the \highecut model produces the lowest
$\chi^2$ value, its cutoff energy at $\sim$3\,keV is so low that this
model effectively turns into a \cutoffpl model. Discrepancies only
occur at very soft energies, where absorption plays a dominant role.
Therefore, different continuum models result in slightly different
$N_\mathrm{H}$ values (see Table~\ref{tab:bestfit_paramer}).

This behavior is very different for Obs~II
(Fig.~\ref{fig:spectrum_ave_obs2}). In particular, the cutoff sets in
much more abruptly than in Obs.~I. Over a wide range below the cutoff,
the spectrum is almost a pure powerlaw, which bends down around
20--30\,keV, although the exact shape of the cutoff is difficult to
disentangle from the broad CRSF. The \cutoffpl model produces strong
residuals at $\sim$30\,keV. These residuals also appear when using the
\fdcut and \npex models, although they are less dominant there. Only
the \highecut model describes the data satisfactorily. Alternatively,
the sharp turnover can also be modeled by a smooth continuum such as
the \cutoffpl or \fdcut and an additional broad Gaussian emission
component which is also shown in Fig.~\ref{fig:spectrum_ave_obs2}. 

We favor the \cutoffpl + \gauss model for Obs.~II for further analysis
because the \npex + \gauss and \fdcut + \gauss models produce CRSF
parameters which indicate that the cutoff is partly modeled by the
CRSF. These models are particularly prone to this problem since they
include both an emission and absorption component which are very close
in energy. The \highecut model does not suffer from this disadvantage
but behaves fundamentally different than the other continuum models
which makes its result difficult to compare to previous work.

We assessed the significance of the inclusion of the Gaussian
component again using the Monte Carlo approach as described for the
iron line. This time, we ran 100\,000 simulations for the \cutoffpl
model and did not find any false-positive detection. The largest
simulated $\chi^2$ difference was $29.0$ which is still very small
compared to the observed one of ${\sim}200$. The nominal significance
of the feature based on our simulation is therefore $>99.999\%$. If we
calculate, however, the probability of a false-positive detection from
the $\chi^2$-distribution for three degrees of freedom, similar to the
approach of \citet{Bhalerao2015}, we obtain a significance of
$>5\sigma$.

\begin{figure}
\centering
\resizebox{\hsize}{!}{\includegraphics{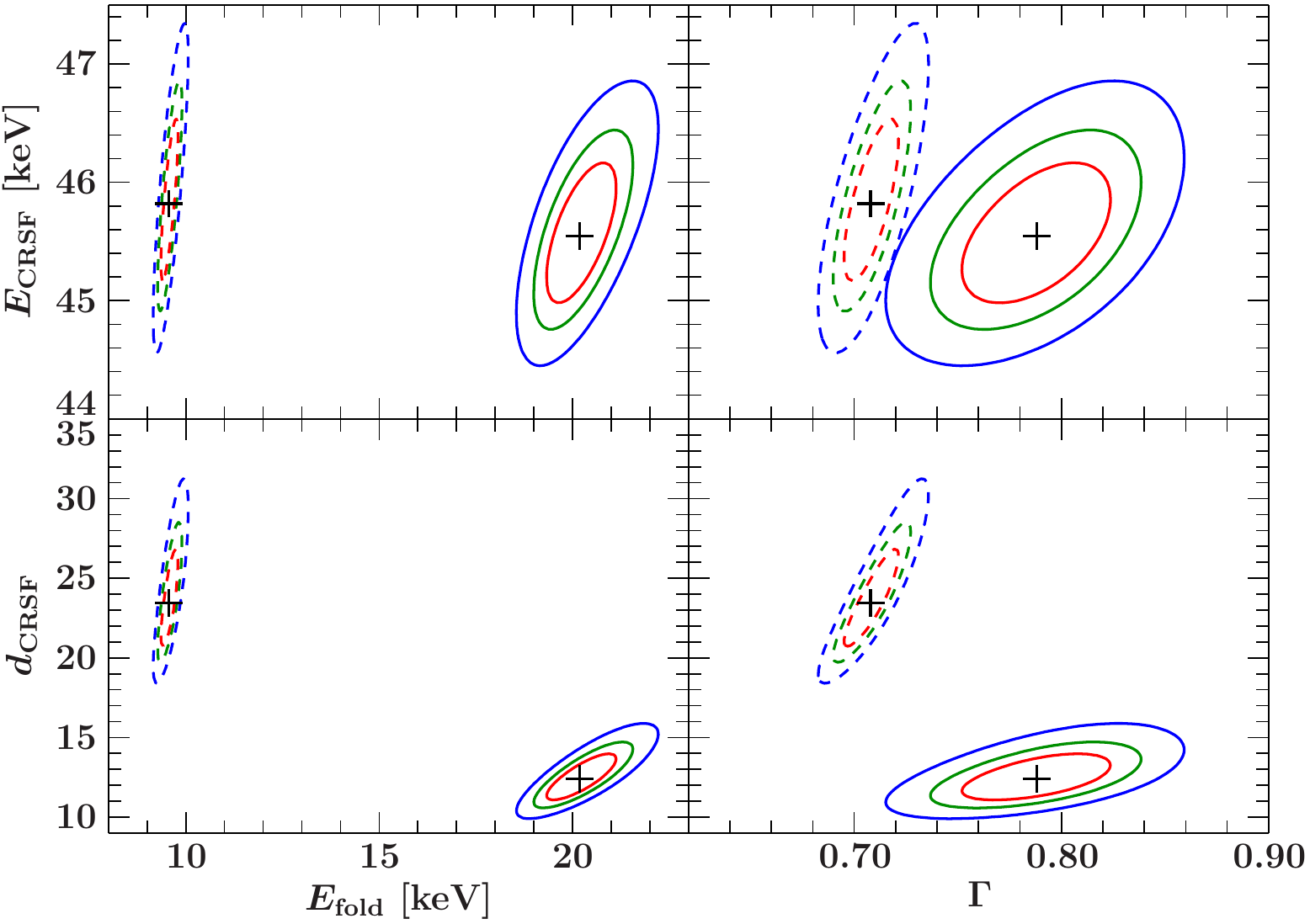}}
\caption{Confidence contours for different fit parameters for Obs.~I
  for the \cutoffpl (solid lines) and for the \npex model (dashed lines). The
  colors red, green, and blue represent 68\%, 90\% and 99\% confidence
  contours, respectively.}
\label{fig:obs1_contours}
\end{figure}

\begin{figure}
\centering
\resizebox{\hsize}{!}{\includegraphics{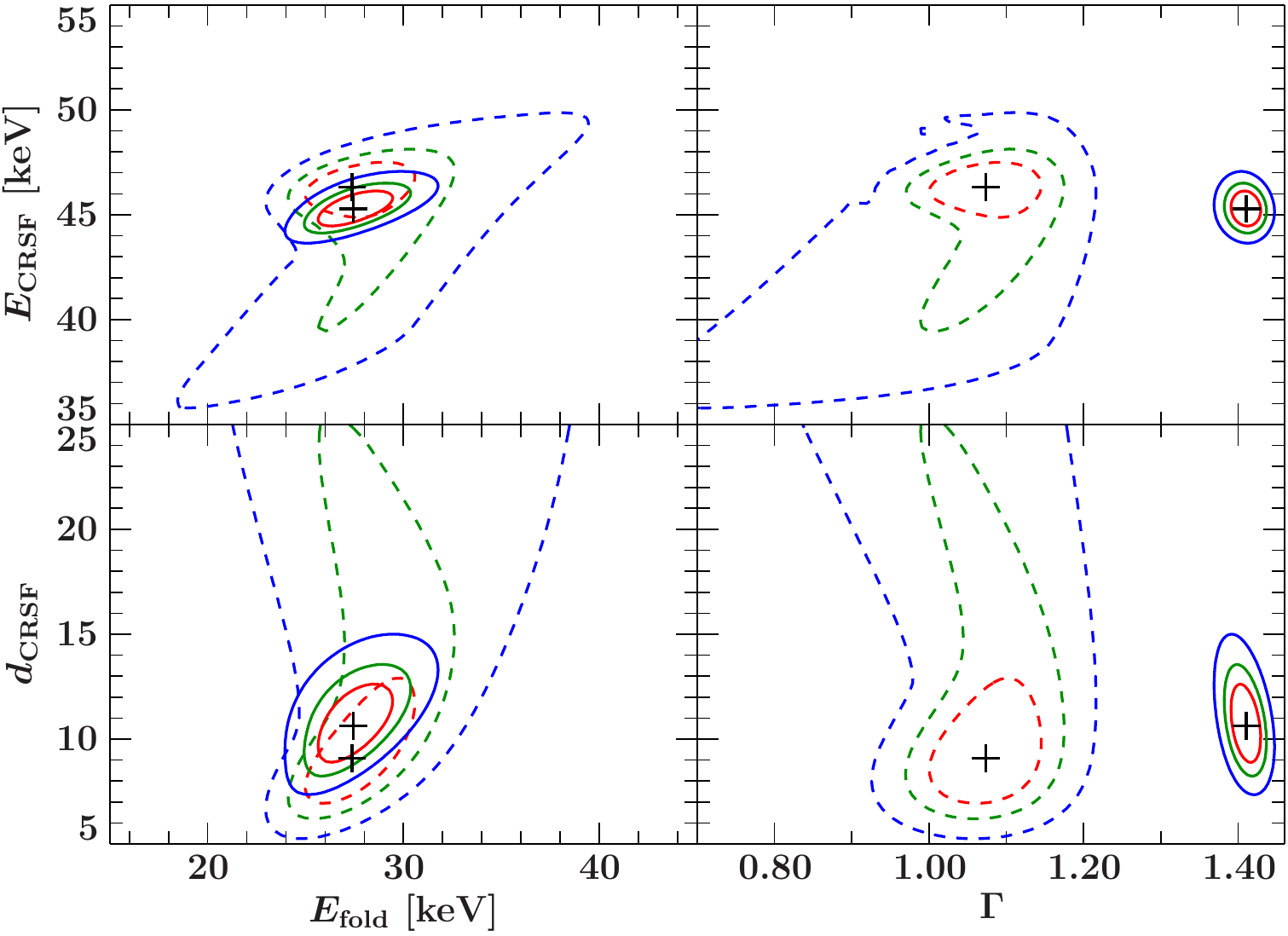}}
\caption{Confidence contours for different fit parameters for Obs.~II
  for the \highecut (solid lines) and for the \cutoffpl + \gauss model
  (dashed lines). The colors red, green, and blue represent 68\%, 90\%
  and 99\% confidence contours, respectively.}
\label{fig:obs2_contours}
\end{figure}

We find that the CRSF energy only depends marginally on the choice of
the continuum model. To ensure that possible artificial correlations
between the CRSF and continuum parameters do not affect our
conclusions about the CRSF, we calculated confidence contours for
several pairs of CRSF and continuum parameters for Obs.~I and II,
respectively, and also compared the different continuum models. The
confidence contours for Obs.~I and II are shown in
Fig.~\ref{fig:obs1_contours} and~\ref{fig:obs2_contours},
respectively. In Obs.~I, higher folding energies correspond to a
stronger CRSF at higher energies, hinting that the exponential
roll-over can partly be modeled by the CRSF. However, this correlation
is very weak. The CRSF energy is very similar for the \cutoffpl and
\npex fits. The different line depths are not unexpected since the
photon indices and the folding energies are very different for the two
models. Observation~II shows well-constrained CRSF parameters for the
\highecut model, and energy and depth of the CRSF are similar for both
the \highecut and the \cutoffpl + \gauss model, although the latter
shows an artificial correlation between the continuum parameters
photon index and folding energy and the CRSF energy and depth. This
behavior is caused by the model containing absorption and emission
components very close to each other in energy.

Similar to the empirical continuum models other than \highecut, the
simple \compmag model can only provide a successful description of the
first, brighter, observation. The second observation shows again clear
residuals around 20--30\,keV (see Fig.~\ref{fig:spectrum_ave_obs2}).
The fit is not acceptable in terms of reduced $\chi^2$ and the
parameters do not settle at physically meaningful values (e.g., we
find accretion column radii that are larger than the canonical neutron
star radius, see Table~\ref{tab:compmag_par}). To produce the best
fit, we set the flag for the velocity profile to be linear in the
optical depth. This is the same assumption as made by
\citep{Becker2007} to analytically solve the radiative transfer
equation. Similar to the behavior of some of the empirical models, an
additional Gaussian emission component at ${\sim}25$\,keV also
improved the fit of the \compmag model to Obs.~II. The best-fit
parameters are also listed in Table~\ref{tab:compmag_par}.

\begin{table}
  \centering
  \caption{Best-fit parameters for \compmag model.}
  \label{tab:compmag_par}
  \begin{tabular}{lccc}
    \hline\hline
     & Obs.~I  & Obs.~II & Obs.~II \\
     & \compmag & \compmag & \compmag+\gauss  \\
    \hline 
    $N_\mathrm{H}$\tablefootmark{a}                 & $0.38^{+0.06}_{-0.05}$           & $0.13^{+0.13}_{-0.12}$            & $0.45^{+0.05}_{-0.06}$\\
    $kT_\mathrm{BB}$\tablefootmark{b}               & $0.97^{+0.02}_{-0.01}$           & $1.08\pm0.02$                     & $1.03^{+0.04}_{-0.03}$ \\
    $kT_\mathrm{e}$\tablefootmark{b}                & $4.35^{+0.16}_{-0.49}$           & $9.93^{+0.01}_{-0.33}$            & $3.7^{+2.0}_{-1.2}$ \\
    $\tau$                                          & $0.75\pm0.04$                    & $0.50\pm0.01$                     & $0.56^{+0.01}_{-0.04}$ \\
    $A$                                             & $0.01$                           & $0.01$                            & $0.01$ \\
    $r_0$\tablefootmark{c}                          & $1726^{+124}_{-223}$             & $>7847$                           & $907^{+537}_{-166}$ \\
    $\mathcal{F}_\mathrm{compmag}$\tablefootmark{d} & $2.96\pm0.01 $                   & $1.03\pm0.01$                     & $0.98^{+0.02}_{-0.07}$ \\
    $\mathcal{F}_\mathrm{Gauss}$\tablefootmark{d}   & --                               & --                                & $0.06^{+0.03}_{-0.02}$ \\
    $E_\mathrm{Gauss}$\tablefootmark{b}             & --                               & --                                & $25.8^{+1.1}_{-0.6}$ \\
    $\sigma_\mathrm{Gauss}$\tablefootmark{b}        & --                               & --                                & $5.6^{+1.1}_{-1.0}$ \\ 
    $E_\mathrm{CRSF}$\tablefootmark{b}              & $45.5\pm0.7$                     & $46.5^{+0.7}_{-0.8}$              & $46.2^{+1.5}_{-2.8}$ \\
    $\sigma_\mathrm{CRSF}$\tablefootmark{b}         & $7.9^{+0.7}_{-0.6}$              & $7.3^{+0.5}_{-0.6}$               & $7.5^{+1.8}_{-1.3}$ \\
    $d_\mathrm{CRSF}$                               & $12.2^{+1.8}_{-1.5}$             & $14.9^{+1.4}_{-1.2}$              & $8.4^{+3.4}_{-1.7}$ \\
    $E_\mathrm{Fe}$\tablefootmark{b}                & $6.38\pm0.05$                    & $6.43^{+0.05}_{-0.12}$            & $6.44^{+0.05}_{-0.12}$ \\
    $\mathrm{EQW}_\mathrm{Fe}$\tablefootmark{e}     & $23\pm4$                         & $15\pm6$                          & $17\pm6$ \\
    $c_\mathrm{FMPA}$\tablefootmark{f}              & $1$                              & $1$                               & $1$ \\
    $c_\mathrm{FMPB}$\tablefootmark{f}              & $1.005\pm0.003$                  & $1.018\pm0.005$                   & $1.018\pm0.004$  \\
    $c_\mathrm{XRT}$\tablefootmark{f}               & $1.129\pm0.016$                  & $0.554\pm0.022$                   & $0.545\pm0.022$ \\
    \hline 
    $\chi^2_\mathrm{red}$ (d.o.f)                   & 1.06 (926)                       & 1.12 (807)                        & 0.96 (804) \\
    \hline 
  \end{tabular}
  \tablefoot{
    \tablefoottext{a}{Equivalent hydrogen column density in units of $10^{22}\,\mathrm{cm}^{-2}$.}
    \tablefoottext{b}{In keV.}
    \tablefoottext{c}{In units of m. Converted from units of Schwarzschild radii for a neutron star mass of $1.4\,M_{\sun}$.} 
    \tablefoottext{d}{3--50\,keV flux in units of $10^{-9}\,\mathrm{erg}\,\mathrm{s}^{-1}\,\mathrm{cm}^{-2}$.}
    \tablefoottext{e}{Equivalent width in units of eV.}
    \tablefoottext{f}{Detector cross-calibration constant with respect to FPMA.}

  }
\end{table}

As an alternative, we tried to model the spectrum of Obs.~II by adding
a second Gaussian absorption line at a lower energy than the
fundamental CRSF. This approach produces statistically acceptable fits
with a $\chi^2_\mathrm{red.}$ of $0.97$ for 804 degrees of
freedom. However, the line energy of the low-energy absorption line is
$15.3^{+1.3}_{-3.5}$\,keV with the \cutoffpl model, making it highly
unlikely that this feature could be the true fundamental CRSF since no
other harmonics than at ${\sim}45$\,keV and ${\sim}100$\,keV have ever
been observed. Furthermore, the line width of $9.8^{+6.6}_{-3.0}$\,keV
is very wide for a low energy CRSF and indicates a continuum modeling
rather than a true absorption feature. We therefore discard the
possibility of a fundamental CRSF at ${\sim}15$\,keV.

\begin{figure}
  \centering
  \resizebox{\hsize}{!}{\includegraphics{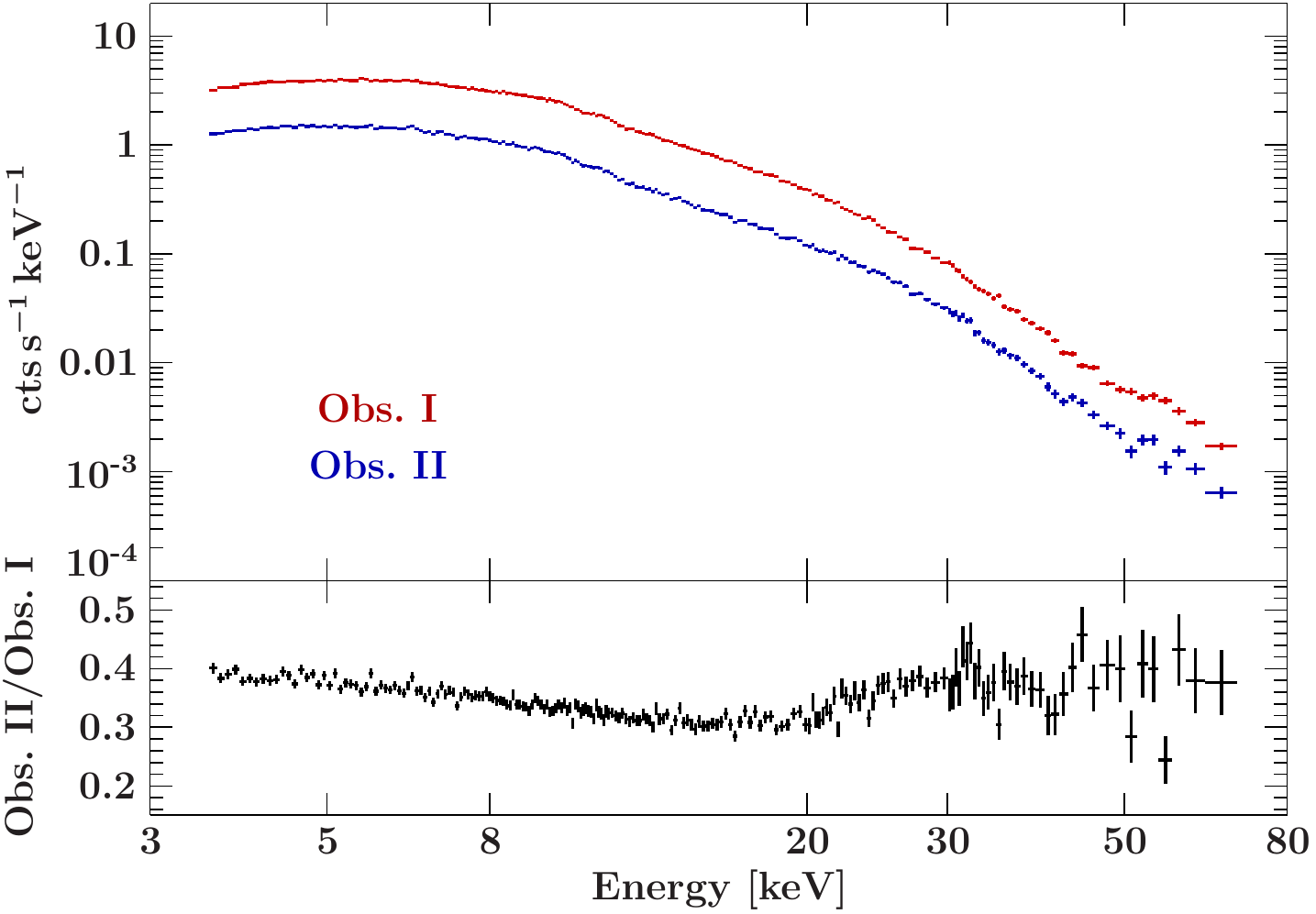}}
  \caption{\textsl{Top:} Count rate spectra of \nustar-FPMA of Obs.~I
    (red) and Obs.~II (blue). \textsl{Bottom:} Ratio of the count rate
    spectra. For clarity we binned the spectra using larger bins for
    the plot than the ones used for the fit.}
\label{fig:spec_ratios}
\end{figure}

For direct comparison of the spectral shape of the two observations,
in Fig.~\ref{fig:spec_ratios} we show the count rate spectra of
\nustar-FPMA and the photon flux ratio. Observation II is softer below
$\sim$20\,keV, but then hardens before the ratio flattens toward
higher energies. This behavior is difficult to track beyond
$\sim$40\,keV, because of the low SNR. The change happens around the
$E_\mathrm{cut}$ energy in the \highecut model and the energy of the
additional Gaussian in the \cutoffpl+\gauss model. The observed
softening between Obs.~I and~II seen in Fig.~\ref{fig:lc_hardness}
reflects the excess of soft photons below 10\,keV.

\section{Pulse phase-resolved
  spectroscopy}\label{sect:spectroscopy_phres}

In order to investigate the variation of the spectral shape with the
viewing angle onto the neutron star, we extracted spectra for 12 pulse
phase intervals from \nustar. The short exposure of the \swift/XRT
observations did not allow to split them further. All pulse
phase-resolved spectra of both observations have been rebinned with
the same requirements as for the pulse phase-averaged spectra and were
restricted to the same energy range.

We fixed the absorption column density to the best fit value obtained
from the phase-averaged spectroscopy with \cutoffpl for Obs.~I and
\cutoffpl + \gauss for Obs.~II. For all continuum models, the energy
of the Fe K$\alpha$ line is consistent with 6.4\,keV and was therefore
fixed to that value to reduce the number of free parameters. We kept
the line narrow again, fixing its width to $10^{-6}$\,keV for both
observations and all phase bins.

As a result of the lower statistics of the pulse phase-resolved
spectra, not all continuum and CRSF parameters can be constrained
simultaneously. Generally, we expect variations of all CRSF parameters
over pulse phase. The centroid energy of the CRSF may depend on the
viewing angle onto the neutron star in a geometrical dipole model
\citep[see, e.g.,][]{Suchy2012} or be Doppler shifted due to viewing
angle-dependent components of the bulk motion of the plasma. The pulse
phase dependence of the width and depth of the CRSF are among other
effects a result of the angle dependence of the scattering cross
sections \citep[see, e.g.,][]{Schwarm2016a}, the plasma temperature,
and the emission pattern of the continuum photons. Since preliminary
fits showed the CRSF energy to be independent of pulse phase, in our
final fits we kept the CSRF energy constant over all phase bins. We
caution, however, that fixing of CRSF or continuum parameters might
introduce artifacts to the spectral modeling. We also note that
\citet{Maitra2013} observed some variability of the CRSF energy over
pulse phase but had to freeze the CRSF width. We fit all the pulse
phase-resolved spectra simultaneously and refer to the parameters that
are constant in all phases as ``global parameters'' \citep[see][for a
description of the method]{Kuehnel2015, Kuehnel2016a}.

Although we expect some variation of the CRSF width over pulse phase,
as observed in Obs.~I, the width of the CRSF could not be constrained
in the pulse phase-resolved spectra of Obs.~II because of the lower
signal-to-noise ratio. We therefore fixed the width of the CRSF to the
value obtained from the phase-averaged fit since our preliminary fits
produced a very wide CRSF. Additionally, we kept the folding energy a
global parameter, because it was not well constrained in all phase
bins, especially the dim phases.

\begin{figure}
  \centering
  \resizebox{\hsize}{!}{\includegraphics{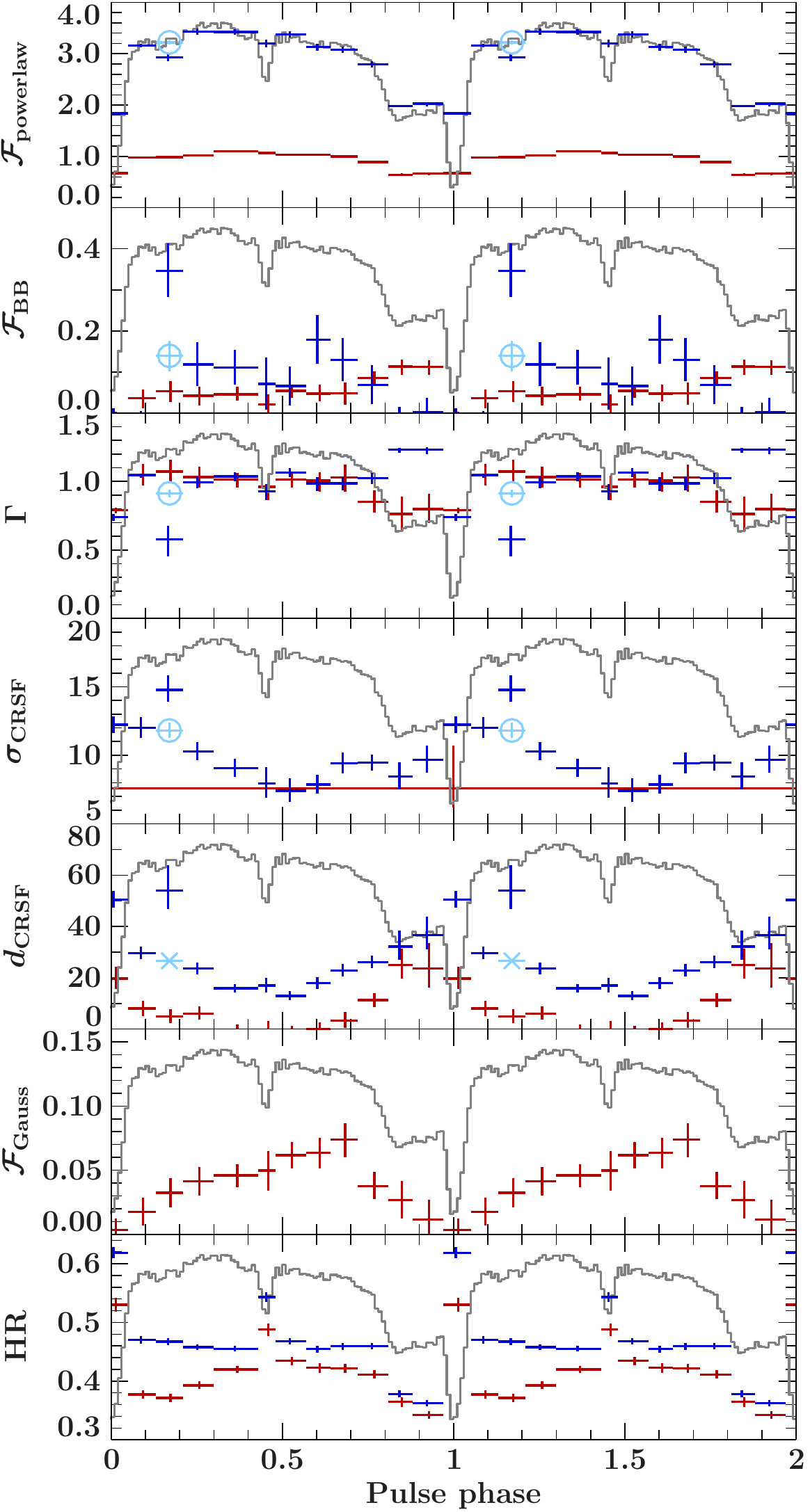}}
  \caption{Evolution of some fit parameters over pulse phase for
    Obs.~I (blue) and Obs.~II (red). The models used were \cutoffpl
    for Obs.~I and \cutoffpl + \gauss for Obs.~II. The width of the
    CSRF could not be constrained for individual pulse phases for
    Obs.~II and was therefore set global. The last panel shows the
    hardness ratio of the individual phase-resolved spectra, defined
    as the count rate ratio of the 15--50\,keV band divided by the
    4--7\,keV band. The gray curve shows the 3--78\,keV pulse profile
    of Obs.~I to illustrate the selection of the phase intervals.
    Pulse profile and parameter evolution are shown twice for clarity.
    All fluxes are given in units of
    $10^{-9}\,\mathrm{erg}\,\mathrm{s}^{-1}\,\mathrm{cm}^{-2}$ and the
    cyclotron line width $\sigma_\mathrm{CRSF}$ in keV. The light blue
    data points in phase interval 0.13--0.21 show an alternative fit
    where the depth of the CRSF was fixed to the mean value of the two
    neighboring bins but all global parameters were kept the same.}
  \label{fig:parameter_evolution}
\end{figure}

For Obs.~II, the energy and width of the Gaussian emission component
only varies marginally with pulse phase, so we kept these parameters
global as well. All other parameters were fitted individually for each
phase interval. The resulting parameter evolutions are shown in
Fig.~\ref{fig:parameter_evolution}. Reduced $\chi^2$ values for the
individual phase intervals ranged from 0.95 to 1.22 for Obs.~I and
0.94 to 1.38 for Obs.~II.

The analysis of the phase-resolved spectra is more prone to
correlations between model parameters than that of the phase-averaged
spectra. This effect is most apparent around phase $\sim$0.17, where
the photon index drops and the black body flux peaks although the
hardness ratio stays rather constant. Fixing the depth of the CRSF for
this particular phase interval to the mean value of the neighboring
bins, however, the CRSF width and the continuum parameters settle at
values favoring the overall evolution of the parameters with pulse
phase (Fig.~\ref{fig:parameter_evolution}, light blue data point).

We obtained the following global parameters: the FPMB normalization
constants are $1.004\pm0.003$ and $1.017\pm0.004$, the black body
temperatures are $1.69\pm0.03$\,keV and $1.30\pm0.02$\,keV, the
folding energies are $32.1\pm0.3$\,keV and $22.7\pm0.2$\,keV, and the
CRSF energies are $47.5\pm0.3$\,keV and $43.1\pm0.5$\,keV for Obs.~I
and II, respectively. Additional global parameters for Obs.~II are the
center and width of the high energy Gaussian, which are
$27.3\pm0.4$\,keV and $5.2\pm0.3$\,keV, respectively.

\section{Discussion}\label{sec:discussion}

\subsection{Timing Analysis}

The binary corrected pulse period only changes marginally between the
two \nustar observations. This is to be expected since at this phase
of the outburst, the mass accretion rate, and therefore the transfer
of angular momentum, are very low. Furthermore, the two observations
are only $\sim$3\,days apart, so intrinsic spin-down should be
negligible.

The morphology of the pulse profiles at low energies as well as their
energy dependence is very similar to the ones observed during the
decay of the 2009 double-peaked outburst when the 3--50\,keV
luminosity was ${\sim}1.2\times10^{36}\,\mathrm{erg}\,\mathrm{s}^{-1}$,
which is close to that of the first \nustar observation
\citep{Caballero2013}.

The pulse profile has been analyzed with a decomposition technique by
\citet{Caballero2011}. They find that the X-ray pulse profiles are
best explained by a hollow accretion column and scattering in a halo
around the polar cap.

\subsection{Continuum variation and modeling}

The pulse phase-averaged spectrum changes significantly between Obs.~I
and~II. While the first, brighter observation can be well described
with common empirical continuum models and the \compmag model, the
spectrum of Obs.~II shows a very sharp cutoff around 30\,keV that
cannot be modeled with most standard continuum models. The sharp
turnover can be modeled by the \highecut model, exploiting its abrupt
steepening at the cutoff energy $E_\mathrm{cut}$. Alternatively, the
``kink'' can also be modeled by introducing an additional Gaussian
emission component. The Gaussian emission component introduces an
additional parameter but results in slightly better fits than the
\highecut model.

The spectrum of \ax could be well described by powerlaw-models with
\emph{smooth} exponential cutoffs over a wide range of luminosities
\citep[e.g.,][ for a 3--50\,keV luminosity range of
$0.04-0.9\times10^{37}\,\mathrm{erg}\,\mathrm{s}^{-1}$]{Caballero2007,Caballero2013}.
\citet{Sartore2015} found that the \cutoffpl model still also
describes \integral observations of \ax at an estimated bolometric
luminosity of ${\sim}4.9\times10^{37}\,\mathrm{erg}\,\mathrm{s}^{-1}$.
Low luminosity and quiescence\footnote{These authors refer to
  luminosities of the order of
  $10^{33}\,\mathrm{erg}\,\mathrm{s}^{-1}$.} observations were taken
with \rxte in 1998 and 2011 \citep{Negueruela2000, Rothschild2013},
with \bepposax in 2000 and 2001 \citep{Orlandini2004}, and \suz
observed \ax in 2005 at a 3--50\,keV luminositiy of
${\sim}3.7\times10^{35}\,\mathrm{erg}\,\mathrm{s}^{-1}$
\citep{Terada2006}. \citet{Terada2006} successfully used an \npex
continuum model while \citet{Orlandini2004} and \citet{Rothschild2013}
found a pure powerlaw model and a thermal bremsstrahlung model to
provide a successful description of the \rxte
data. \citeauthor{Rothschild2013} caution, however, that the
non-detection of an exponential cutoff could be due to the low
signal-to-noise ratio. \suz also observed \ax in 2009 at a 3--50\,keV
luminosity of ${\sim}4\times10^{35}\,\mathrm{erg}\,\mathrm{s}^{-1}$
\citep{Caballero2013}, at a brightness very similar to the dimmer
\nustar observation. The phase-averaged spectrum of this observation
were modeled with a \cutoffpl model \citep{Caballero2013} and a
partially covered \npex, powerlaw and \texttt{compTT} model
\citep{Maitra2013}. All these fits showed moderate residuals near
30\,keV that are similar to those of our fits of Obs.~II with the
\cutoffpl models \citep[][Fig.~3 and 4, respectively]{Caballero2013,
  Maitra2013}. The two \nustar observations presented here are
slightly brighter than the 2005 \suz observation and cover the
luminosity of the dimmest \rxte observations, but provide higher
sensitivity compared to \suz/PIN and \rxte/PCA. It is therefore very
likely that this spectral change would have been unobserved, even if
it had happened in previous outbursts.

The luminosity estimates quoted above all used a distance of 2\,kpc.
This value was derived from spectroscopic measurements
\citep[e.g.][]{Hutchings1978, Giangrande1980, Steele1998}.
\citet{Giangrande1980} report an uncertainty of the spectroscopic
distance measurement of 0.6\,kpc which introduces a systematic
uncertainty of the luminosity of a factor of $\sim$4. Further
systematic uncertainties are introduced, e.g., by assuming isotropic
emssion and are discussed in detail in \citet{Martinez-Nunez2017},
\citet{Kuehnel2016b}, and Falkner et al.\ (in prep.).

\begin{figure}
  \centering
  \resizebox{\hsize}{!}{\includegraphics{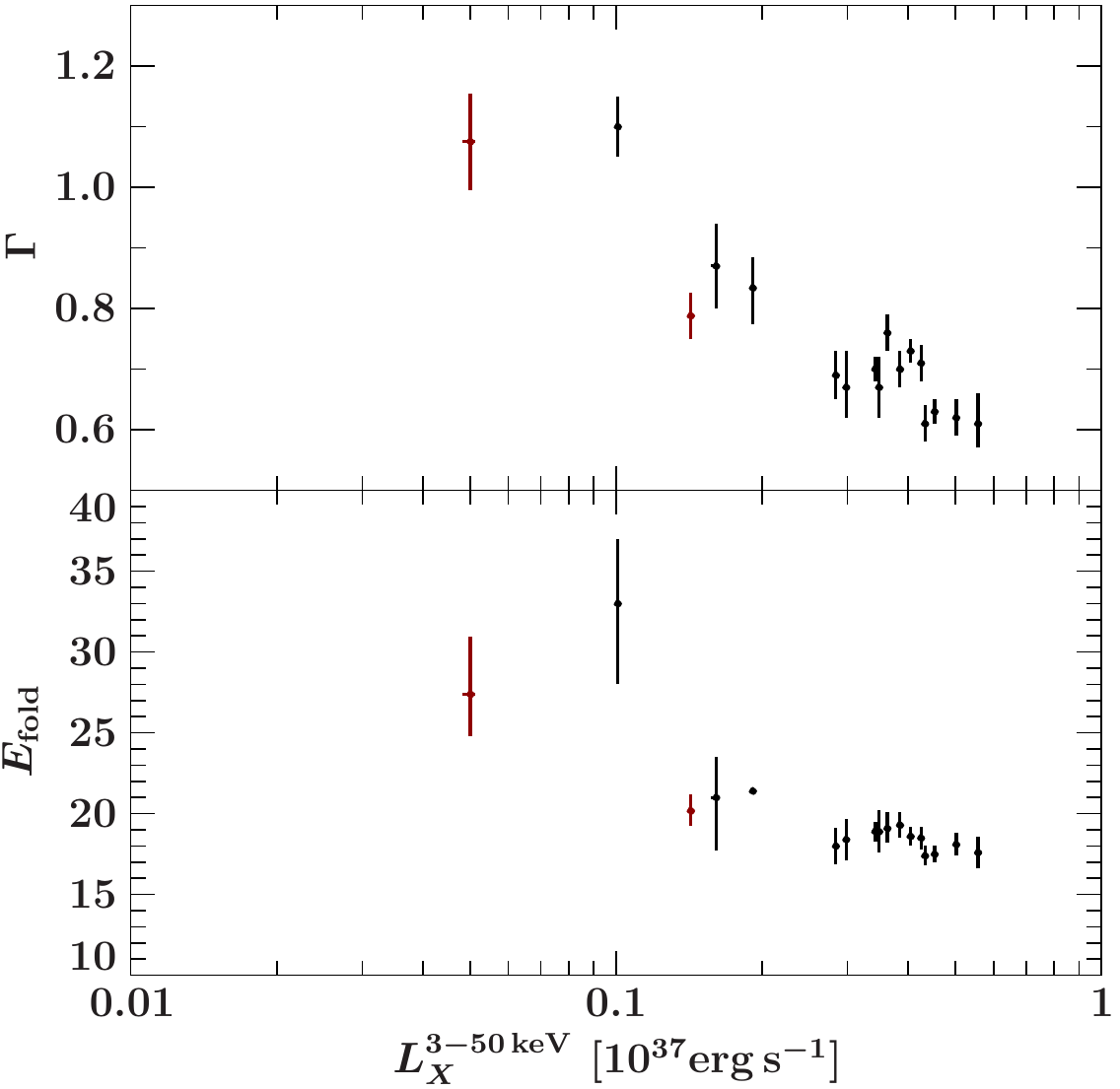}}
  \caption{Evolution of the continuum parameters $\Gamma$ and
    $E_\mathrm{fold}$ with the 3--50\,keV luminosity. Black data
    points show the \rxte results of \citet{Caballero2013}, red data
    points show the result of the \nustar observations. The former
    were obtained using a \cutoffpl continuum model, the \nustar
    results using a \cutoffpl and a \cutoffpl+ \gauss model for Obs.~I
    and II, respectively.}
  \label{fig:continuum_parameter}
\end{figure}

For further comparison, we show the evolution of the continuum
parameters $\Gamma$ and $E_\mathrm{fold}$ with the 3--50\,keV
luminosity of the \nustar observations and the \rxte observations of
\citet{Caballero2013} in Fig.~\ref{fig:continuum_parameter}. The \rxte
photon index increases toward lower luminosities, which is nicely
confirmed with \nustar. The folding energy observed by \rxte is rather
constant for luminosities of
$\sim(3\mbox{--}6)\times10^{36}\,\mathrm{erg}\,\mathrm{s}^{-1}$ and
then increases with decreasing luminosity. While the \nustar
observations are in line with this behavior, they indicate a weaker
correlation of the folding energy with luminosity. We note, however,
that when fitting the \cutoffpl model without the Gaussian emission
component to Obs.~II, the photon index is harder with similar folding
energies. \citet{Caballero2013} also report on a \suz observation at a
luminosity comparable to the fainter \nustar observation. These
authors used again a \cutoffpl model and found a photon index around
${\sim}0.84$. The fit of \nustar Obs.~II is significantly improved by
adding the Gaussian component (the $\chi^2$ changes from 974.9 to
773.3 for the \cutoffpl model) and the photon index reflects the
softening also shown in Fig.~\ref{fig:lc_hardness}.

This overall softening is in line with earlier work on the spectral
behavior of \ax at lower luminosities. Using \rxte and \integral data
collected in 2010 April at 10--100\,keV luminosities of
$\sim(0.1\mbox{--}1.2)\times10^{37}\,\mathrm{erg}\,\mathrm{s}^{-1}$ in
10--100\,keV, \citet{Mueller2013b} observed a softening of the photon
index and an increase of the folding energy toward lower luminosities.
This result confirmed earlier work by \citet{Klochkov2011}, who report
a spectral softening for decreasing luminosity in their
pulse-height-resolved spectroscopy at a luminosity level of around
$10^{38}\,\mathrm{erg}\,\mathrm{s}^{-1}$. Due to lower signal to noise
ratios, \citet{Klochkov2011} had to fix the folding energy to
constrain the $\Gamma$-luminosity correlation.

The physical explanation of this softening at low luminosities is not
clear. Whiles pectral formation in accretion columns is an area of
active study that started with \citet{Basko1975} and is under
continued refinement \citep[see, e.g.,][]{Becker2007, Postnov2015},
few authors focus on spectral formation at very low luminosities.
\citet{Postnov2015} observed a softening of the X-ray spectrum toward
lower luminosities in a sample of six accreting pulsars and compared
this observational result to numerical calculations of the
two-dimensional structure of the accretion column with a radiation
dominated shock. They focussed on luminosities of
$10^{37}\,\mathrm{erg}\,\mathrm{s}^{-1}$ and above and found that the
behavior of the spectral hardness is reproduced by Compton saturated
emission from an optically thick accretion column. A saturation of the
hardness ratio at a few times $10^{37}\,\mathrm{erg}\,\mathrm{s}^{-1}$
was observed and explained by \citeauthor{Postnov2015} as reflection
from the neutron star surface. Their calculations, which are based on
the results by \citet{Lyubarskii1986}, however are only valid for
photon energies far below the CRSF energy.

At lower luminosities, \citet{Langer1982} discuss accretion onto
highly magnetized ($B\sim10^{12}\,\mathrm{G}$) neutron stars for
accretion rates below $10^{16}\,\mathrm{g}\,\mathrm{s}^{-1}$. In this
accretion regime, a collision-less shock forms and radiation braking
becomes negligible. \citeauthor{Langer1982} assume that the spectral
distortion of the seed photons due to Comptonization can be neglected.
The assumptions for their model are fulfilled by \ax at the luminosity
level of Obs.~II \citep[][Eq.~25, where we used the accretion column
  radii of Table~\ref{tab:compmag_par}]{Langer1982}. Interestingly,
the sample spectrum shown by \citet[][their Fig.~4]{Langer1982} is
based on system properties ($B\sim5\times10^{12}\,\mathrm{G}$ and
$\dot{M}\sim5\times10^{15}\,\mathrm{g}\,\mathrm{s}^{-1}$) which are
very close to those of Obs.~II. The predicted spectral shape, however,
is clearly different from that observation. The authors argue that
most of the energy is emitted in a Doppler-broadened cyclotron
emission component, which does not represent the overall powerlaw-like
shape we observe. If the accretion rate decreases even further, the
height of the shock is expected to increase and while the emission is
still dominated by cyclotron emission, it originates over a wider
range along the column. This results in a superposition of different
cyclotron energies and forms a smooth continuum with an exponential
cutoff above the CRSF energy at the surface. Comparing their model to
the \nustar observation leads, however, reveals significant problems
in this interpretation: The observed cutoff energy is below the
measured CRSF energy and the spectral shape in the model deviates
significantly from a powerlaw below the cutoff energy. 

A newer model for the emission of accreting neutron stars at low
luminosities has recently been discussed by \citet{Vybornov2017}, who
studied the behavior of \object{Cep X-4} at luminosities of
${\sim}1.5\times10^{36}\,\mathrm{erg}\,\mathrm{s}^{-1}$ and
${\sim}6\times10^{36}\,\mathrm{erg}\,\mathrm{s}^{-1}$, i.e.,
comparable to those the luminosity range studied here for \ax.
\citeauthor{Vybornov2017} show that the behavior of \object{Cep X-4}
at low luminosities can be described using the combination of a
collisionless shock and unsaturated Comptonization, rather than
undistorted cyclotron emission \citep{Langer1982}. In this
Comptonization picture, the spectral shape is powerlaw-like below the
CRSF energy and softening toward low luminosities
\citep{Vybornov2017}. \citet{Vybornov2017} show that the evolution of
the hardness in \object{Cep X-4}, which also shows softening with
luminosity, is consistent with this shock picture. Unfortunately, the
spectral model used by \citet{Vybornov2017} is not directly applicable
to the broad band data used here. By looking at the Compton-$y$
parameter, however, we can test whether the spectral fits found for
the \compmag model are in the same parameter regime as that claimed
for \object{Cep X-4}. For a strong magnetic field, the Compton
$y$-parameter is given by \citep{Basko1975}
\begin{equation}
y =\frac{2}{15}\frac{kT_\mathrm{e}}{m_\mathrm{e}c^2}\max(\tau, \tau^2)~,
\end{equation}
with optical depth, $\tau$, and electron temperature, $kT_\mathrm{e}$.
Based on the hardness evolution of \object{Cep X-4},
\citet{Vybornov2017} find $y$ to range between $y=0.2$ and $y=1.2$. In
contrast, using the \compmag parameters from
Table~\ref{tab:compmag_par}, we find
$y\sim(5\mbox{--}8)\times10^{-4}$.  Taking these values at face value,
this result implies \object{Cep X-4} and \ax to be in different
accretion regimes, despite their similar luminosities. We note,
however, that the systematic uncertainty of these $y$-values is very
large. Given the general behavior seen, however, further work
extending the spectral model to the energy range considered here would
be very desirable.

Finally, the electron temperatures around found in our fits of the
spectra of \ax are comparable to the values given for example cases
\object{LMC X-4}, \object{Cen X-3}, and \object{Her X-1} by
\citet{Becker2007}, although these authors considered higher
luminosities. It is also close the application of their model to the
spectrum of \object{Her~X-1} \citep{Wolff2016}. \citet{Farinelli2016}
applied an advanced version of the \compmag model to data of
\object{Cen X-3}, \object{4U 0115+63}, and \object{Her X-1} and found
smaller electron temperatures of 0.8--3\,keV which they explained
being due to the inclusion of second order bulk Comptonization in the
RTE. Comparing our \compmag fits to the successful application of the
same version of the model to data of the accreting pulsar
\object{RX~J0440.9+4431} by \citet{Ferrigno2013}, however, the
resulting Compton-$y$ values are consistently much smaller than unity
in both cases (although we find higher optical depths and lower
electron temperatures than these authors).

Regarding the Gaussian emission component that is required with the
smooth continuum models of the fainter observation, we note that
\citet{Iwakiri2012} observed a Gaussian-shaped emission feature during
the dim pulse phase of \object{4U 1626$-$67}. They interpret this
feature as the CRSF which appears in emission only during that
particular pulse phase due to the lower optical depth along the line
of sight for the corresponding viewing angle. During all other phases
and in the phase-averaged spectrum, the CRSF clearly appears in
absorption.

This behavior is clearly different from our fainter observation of
\ax. The Gaussian emission component is clearly visible in the
phase-averaged spectrum, while a CRSF in absorption is still
required. In the pulse phase-resolved spectroscopy, the Gaussian
emission component is faintest during the dim phases, which is the
opposite behavior of what \citet{Iwakiri2012} found.

\subsection{Luminosity dependence of the CRSF}

One goal of our observations was to investigate the CRSF-luminosity
dependence of \ax toward very low luminosities: As the mass accretion
rate changes, we expect changes in the geometry of the accretion
column. Since the CRSF energy is representative of the magnetic field
strength in the region in which most of the radiation is produced, we
expect changes in the column geometry to have an impact on the
measured line energy \citep[][and references therein]{Becker2012}. For
a long time, no such changes were seen: Observations of many different
outbursts over more than two decades revealed the CRSF line energy to
be stable \citep[see, e.g.,][]{Kendziorra1994, Terada2006,
  Caballero2007, Caballero2013}. Indications of a positive correlation
of the CRSF energy with luminosity were reported by
\citet{Klochkov2011} in a pulse-height-resolved analysis of \rxte and
\integral data and by \citet{Sartore2015} using pulse-averaged
spectroscopy. The latter authors also observed significant changes in
the continuum shape but had to fix some of the continuum parameters
due to strong model-intrinsic parameter correlations. This common
approach, however, makes the impact of the continuum variation on the
cyclotron line parameters difficult to estimate. All data showing such
indications of a positive correlation of the CRSF energy with
luminosity were taken at luminosities of
${\sim}10^{37}\,\mathrm{erg}\,\mathrm{s}^{-1}$ and above.

\begin{figure}
\centering
\resizebox{\hsize}{!}{\includegraphics{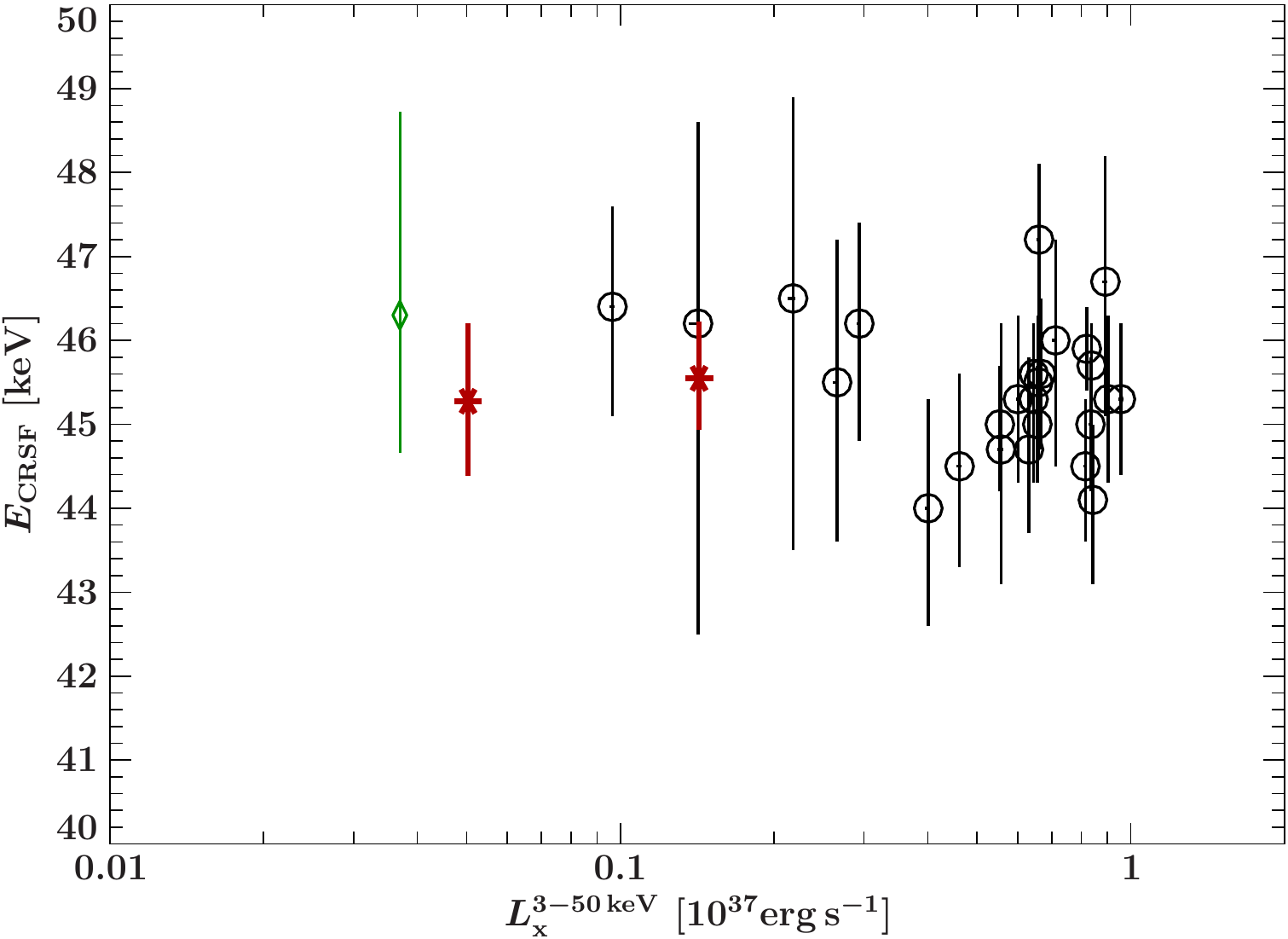}}
\caption{Luminosity dependence of the cyclotron line energy. The black
  circles and green diamond denote \rxte measurements from
  \citet{Caballero2007} and \citet{Terada2006}, respectively, while
  the red crosses show the \nustar results. The continuum models used
  to measure the line energy are \cutoffpl and \highecut for Obs.~I
  and Obs.~II, respectively. For Obs.~II, we used the result of the
  fit with the \highecut model, since, while the \highecut and
  \cutoffpl + \gauss models result in consistent values for
  $E_\mathrm{CRSF}$, this value is better constrained in the \highecut
  case (Fig.~\ref{fig:obs1_contours} and
  \ref{fig:obs2_contours}). \citet{Terada2006} and
  \citet{Caballero2007} used an \npex and \cutoffpl model,
  respectively.}
\label{fig:lx_ecycl}
\end{figure}

Previous CRSF observations at low luminosities (the only observations
taken at a luminosity lower than the one considered here were taken by
\suz in 2005, \citeauthor{Terada2006}, \citeyear{Terada2006} and in
2009,\citeauthor{Caballero2013}, \citeyear{Caballero2013}) and
\citet{Caballero2007}, and \citet{Maitra2013} did not show changes of
the CRSF energy with luminosity. Our \nustar analysis confirms this
result with a much higher precision than was possible with previous
missions (see comparison with \citeauthor{Terada2006},
\citeyear{Terada2006} and \citeauthor{Caballero2007},
\citeyear{Caballero2007} in Fig.~\ref{fig:lx_ecycl}, and Fig.~4 in
\citeauthor{Caballero2013}, \citeyear{Caballero2013}).

The CRSF width is typically around 10\,keV
\citep[e.g.,][]{Caballero2009b, Sartore2015}, which is in good
agreement with our results, although smaller line widths have also
been reported \citep[e.g.,][]{Terada2006}. In contrast, the line depth
has been observed to vary significantly between different
observations. In principle, such a variability of the depth could be
caused by changes in the accretion rate (although the optical depth in
the CRSF core is always very high), but we note that the depth depends
on the choice of the continuum model (e.g., \npex produces deeper
lines, see Table~\ref{tab:bestfit_paramer}), and caution is advised
when comparing data using different continua. At least for the
luminosity range covered by \citet{Sartore2015} and by our \nustar
data, when using the same continuum model the optical depths remain
independent of the luminosity.

\subsection{Pulse phase-resolved spectroscopy}

In the following we discuss the variability of \ax with the pulse
phase. The spectrum is strongly variable in both data sets. Despite
their absolute values being considerably different, the hardness
ratios (see last panel of Fig.~\ref{fig:parameter_evolution}) shows a
similar trend in both observations.

Of special interest are the phase bins around 0.0 and 0.45, where a
large jump in the hardness ratio is seen. While the photon index shows
a similar jump at phase 0.0, it hardly deviates from its average value
at phase 0.45. We note, however, that the latter behavior could also
be due to parameter correlations similar to those discussed in
Sect.~\ref{sect:spectroscopy_phres} for the bin around phase
$\sim$0.17.

The additional Gaussian emission component around 26\,keV in Obs.~II
is also required for the phase-resolved spectral analysis. The
centroid energy and width of this component could not be constrained
for each pulse phase interval individually and were therefore kept
global throughout the fit. The flux of the Gaussian component varies
only moderately over pulse phase (see
Fig.~\ref{fig:parameter_evolution}) and is of the same order of
magnitude as the black body flux.

The overall behavior of the source with pulse phase is in line with
that discussed by \citet{Maitra2013} in their pulse phase-resolved
spectroscopy of \ax at a luminosity of
${\sim}5\times10^{35}\,\mathrm{erg}\,\mathrm{s}^{-1}$. These authors
applied an \npex as well as a \texttt{compTT} continuum in a partial
covering absorption geometry. They found an increase of the covering
fraction and the local absorption component during the main dip of the
pulse profile minimum and associate this with a narrow accretion
stream. Due to \nustar's higher low energy threshold, our spectral
fits did not require a partial covering model, but similar to
\citet{Maitra2013} we find that the hardness ratio has its maximum at
the pulse profile minimum, also implying a lack of soft photons at
this specific pulse phase.

Turning to spectral components at higher energies, CRSF parameters are
expected to vary significantly over pulse phase as observed in many
sources, e.g., in \object{Cen X-3} \citep{Burderi2000} or \object{2S
  1553$-$542} \citep{Tsygankov2016}, so fixing any CRSF parameters can
introduce artifacts into the spectral modeling. Therefore, a
compromise between maximizing signal-to-noise, avoiding parameter
degeneracies and capturing intrinsic spectral variability had to be
found. We find that, in contrast to other analyses
\citep[e.g.,][]{Maisack1997, Maitra2013}, holding the line energy
constant and letting its depth vary results in more consistent results
(the width had to be fixed in Obs.~II and was let free in
Obs.~I). This approach is motivated by the assumption that the pulse
phase variability of the line is mainly due to changes of the
direction of our line of sight onto the CRSF forming region, while
still mainly seeing the same emission region and thus magnetic field.
In general, the \nustar data show that the general behavior of the
CRSF is similar in both observations and the CRSF is deepest around
the pulse profile minimum. This result is in disagreement with
\citet{Maitra2013}, who found the depth of the CRSF to increase with
pulse phase during the main peak and then drop significantly around
the pulse profile minimum, accompanied by a sudden change in observed
CRSF energy. We consider this behavior to be less likely, as our
modeling approach does not introduce sudden changes in CRSF parameters
with phase.

Next, we try to connect the observed CRSF parameter evolution to an
emission geometry. Theoretical calculations and Monte-Carlo
simulations of cyclotron resonant scattering of photons in an electron
plasma predict \citep[][and references therein]{Schwarm2016a} that
CRSF get wider and shallower when the angle of the photons to the
magnetic field in the rest frame of the plasma becomes smaller and
narrower but deeper when the angle to the magnetic field gets
larger. This behavior is a result of the width of the peaks of the
resonances of the scattering cross sections, which are highly
angle-dependent \citep{Schwarm2016a}. This general result is, however,
only strictly true for the higher harmonic lines, since the line depth
of the fundamental is significantly affected by photon spawning, i.e.,
the emission of resonant photons during the successive decay of
electrons from higher Landau levels, which fill up the fundamental
absorption line.

In a very simplified picture we assume pure fan beam emission of the
accretion column where the radiation of one accretion column always
dominates the observed radiation. Generally, the fan beam emission
pattern is supposed to be generated at high luminosities but a precise
luminosity measurement from observational data is very difficult
because of numerous systematic uncertainties \citep{Kuehnel2016b}. The
pulse profile shape in the \nustar observations is similar to those
observed at higher luminosities \citep[e.g.,][]{Caballero2013} so it
is unlikely that the emission geometry has changed much compared to
higher luminosities. The fan beam scenario is further supported by
\citet{Caballero2011} who studied the accretion geometry and emission
pattern of \ax applying a pulse profile decomposition method, although
at luminosities of
${\sim}10^{37}\,\mathrm{erg}\,\mathrm{s}^{-1}$. However, the evaluation
of their model, which includes a hollow accretion column, a scattering
halo above the neutron star surface and a complex beam pattern is
beyond the scope of current CRSF simulations, although more
sophisticated accretion column geometries are very well accessible
with the Monte Carlo approach \citep{Schwarm2016b}.

One possible result of the fan beam emission can be that indeed most
flux is observed at large viewing angles to the $B$-field and,
therefore, that these viewing angles are connected to the pulse
profile maximum. We caution that our proposition of the maximum of the
pulse profile being connected to large viewing angles due to fan beam
emission geometry is a drastic simplification of the highly complex
process of pulse profile formation which includes relativistic effects
such as boosting, gravitational light bending, a non-isotropic
emission pattern as well as geometrical properties of the system
(e.g., inclination of the $B$-field and the observer, height, size,
and position of the accretion column). 

In our observations, the CRSF width has its minimum during the broad
peak of the pulse profile. Assuming small CRSF widths are associated
with large viewing angles, as simulations of CRSF formation mentioned
above indicate, this supports the simplified fan beam scenario. We
find that changes of the CRSF width and depth with pulse phase are
correlated, i.e., the CRSF becomes narrowest and shallowest and vice
versa ( see Fig.~\ref{fig:parameter_evolution}). This disagrees with
the predictions of simulations because the phase dependence of the
width and depth of the CRSF should be determined by the angle to the
magnetic field, with opposite angle dependence for the width and depth
of the CRSF.

One possible explanation for the unexpected behavior of the CRSF depth
could be that the fundamental line is strongly affected by photon
spawning of the harmonic lines (the second harmonic at $\sim$100\,keV
has been observed, e.g., by \citealt{Kretschmar1996},
\citealt{Orlandini2004}, and \citealt{Sartore2015}). Furthermore, the
excitation rates of fundamental and harmonic lines, and consequently
the impact of photon spawning, depend on the hardness of the
underlying continuum, which varies with pulse phase. Additionally,
different viewing angles may correspond to different optical depths of
the column, which strongly determine the observed line
depths. Finally, the observed spectrum is the sum of the spectra from
two accretion columns seen under different angles, which would further
complicate the picture. All these effects could produce an opposite
pulse phase dependence of the fundamental CRSF depth as compared to
simulations. A proof of this conjecture could be provided by pulse
phase-resolved spectroscopy of both the fundamental and the second
harmonic, although the latter is not accessible with \nustar.

\section{Conclusions and Outlook}

In this paper we have reported on two \nustar observations of \ax at
3--50\,keV luminosities of
${\sim}1.4\times10^{36}\,\mathrm{erg}\,\mathrm{s}^{-1}$ and
${\sim}5\times10^{35}\,\mathrm{erg}\,\mathrm{s}^{-1}$, respectively.
These luminosities are comparable to the faintest observations of \ax
taken so far with other instruments \citep{Terada2006}. The quality of
the observations allows for a precise measurement of the CRSF
parameters as well as the continuum. The CRSF energy is in agreement
with previous measurements, confirming that the CRSF energy is
independent of luminosity over a wide range of luminosities. Together
with the constancy of the pulse profile with energy, they are probably
indicating that the accretion column is stable at these luminosities
considered here.

The continuum shape changes significantly between the two
observations. While the first, brighter observation is similar in
spectral shape to more luminous observations, the second, fainter one
shows a ``kink''-like feature around 25\,keV which can be either
modeled by a an abrupt exponential cutoff or an additional emission
component on top of a smooth continuum. There have been indications of
a spectral transition at a luminosity of
0.5--1.0$\times10^{36}\,\mathrm{erg}\,\mathrm{s}^{-1}$ in earlier
observations, e.g. by remaining residuals around 20--30\,keV in the
\suz observation in 2009 \citep{Maitra2013, Caballero2013}, but
\nustar could now resolve this evolution with unprecedented
quality. The spectral softening toward lower luminosities found here
confirms previous observations. The evolution of the photon index and
folding energy is in line with \citet{Caballero2013}. A similar
behavior of the spectral hardness was reported by \citet{Vybornov2017}
for \object{Cep X-4}. These authors could reproduce the spectral
hardness evolution with an unsaturated Comptonization in a
collisionless shock model, however, leading to very different values
of the Compton $y$-parameter.

For a detailed quantitative interpretation of the pulse-phase
dependence of the observed spectra, the geometry of the source has to
be determined and self-consistent, physical spectral models need to be
developed that take the angle- and height-dependence of the emitted
radiation into account. One step toward an understanding of the
geometry of \ax was provided by \citet{Caballero2011} who determined
the energy-resolved emission pattern of \ax with a pulse-profile
decomposition method. A self-consistent accretion column model that
combines continuum and CRSF formation with relativistic effects such
as light bending is currently under development (Falkner et al., in
prep.).

Observations of the CRSF energy-luminosity dependence as well as the
dependence of the continuum shape on the luminosity of accreting
pulsars over a wide range of luminosities are essential for the
further understanding of the structure and formation of the accretion
columns, since the physical conditions inside the column are expected
to depend strongly on the mass accretion rate, which can vary by
several orders of magnitude even for individual sources. From an
observational point of view, nearby sources such as \ax or
\object{GX~304$-$1} are of particular interest because they allow
observations at very low luminosities at sufficient signal-to-noise
ratio with moderate exposure time. The theory of accretion columns
focused mainly on the high-luminosity cases in recent years. We expect
additional high-quality observations of accreting pulsars at very low
luminosities to foster the development of general theoretical models
of low-luminosity accretion on neutron stars.

\begin{acknowledgement}
  We thank Sebastian Falkner, Matthias K\"uhnel, and Ingo Kreykenbohm
  for many fruitful discussions. We thank the Deutsches Zentrum f\"ur
  Luft- und Raumfahrt for support under contract 50\,OR\,1410. This
  work was supported under NASA Contract No.\ NNG08FD60C, and made use
  of data from the \nustar mission, a project led by the California
  Institute of Technology, managed by the Jet Propulsion Laboratory,
  and funded by the National Aeronautics and Space Administration. We
  thank the \nustar Operations, Software and Calibration teams for
  support with the execution and analysis of these observations. This
  research has made use of the \nustar Data Analysis Software
  (NuSTARDAS) jointly developed by the ASI Science Data Center (ASDC,
  Italy) and the California Institute of Technology (USA).

  This work has made use of data from the European Space Agency (ESA)
  mission \textit{Gaia} (\url{http://www.cosmos.esa.int/gaia}),
  processed by the \textit{Gaia} Data Processing and Analysis
  Consortium (DPAC,
  \url{http://www.cosmos.esa.int/web/gaia/dpac/consortium}). Funding
  for the DPAC has been provided by national institutions, in
  particular the institutions participating in the \textit{Gaia}
  Multilateral Agreement.

  This research has made use of a collection of ISIS functions
  (ISISscripts) provided by ECAP/Remeis observatory and MIT
  (\url{http://www.sternwarte.uni-erlangen.de/isis/}).

\end{acknowledgement}

\bibliographystyle{aa}
\bibliography{references}

\end{document}